\documentclass[pdflatex,sn-mathphys-num]{sn-jnl}

\usepackage{graphicx}%
\usepackage{multirow}%
\usepackage{amsmath,amssymb,amsfonts}%
\usepackage{amsthm}%
\usepackage{mathrsfs}%
\usepackage[title]{appendix}%
\usepackage{xcolor}%
\usepackage{textcomp}%
\usepackage{manyfoot}%
\usepackage{booktabs}%
\usepackage{algorithm}%
\usepackage{algorithmicx}%
\usepackage{algpseudocode}%
\usepackage{listings}%
\usepackage{subcaption}

\begin{document}

\title[Article Title]{Wind turbine enhancement via Active Flow Control Implementation}


\author[1]{\fnm{Marc} \sur{Lahoz}}\email{marc.lahoz10@gmail.com}

\author[1]{\fnm{Ahmad} \sur{Nabhani}}\email{ahmad.nabhani@upc.edu}

\author[1]{\fnm{Mohammad} \sur{Saemian}}\email{mohammad.saemian@upc.edu}

\author*[1]{\fnm{Josep M} \sur{Bergada}}\email{josep.m.bergada@upc.edu}

\affil*[1]{\orgdiv{Fluid Mechanics Department,} \orgname{Universitat Politècnica de Catalunya,} \orgaddress{\street{Jordi Girona 31}, \city{} \postcode{08034}, \state{Barcelona}, \country{Spain}}}


\abstract{ The present research aims to enhance the efficiency of a DTU-10MW Horizontal Axis Wind Turbine (HAWT) via Active Flow Control (AFC) implementation and using Synthetic Jets (SJ). In the initial part of the study the flow around two airfoil sections cut along the wing turbine blade and for a wind speed of $10m/s$ is simulated using CFD-2D-RANS-K$\omega$-SST turbulence model, in order to obtain the time averaged boundary layer separation point and the associated vortex shedding frequency. This information is used, on a second stage of the paper, to set, in one of these two airfoils where the boundary layer is having an early separation, two of the AFC parameters while optimizing the remaining three. The optimization is performed employing a parametric analysis and demonstrates that a considerable WT power increase can be obtained when managing to reattach the former separated boundary layer. This is further clarified thanks to the energy assessment presented in the final part of the paper. Although the AFC optimization needs to be extended for the rest of the blade sections, the procedure outlined in the present research clarifies the different steps to be followed to optimize the performance of any HAWT under any operating conditions. The Reynolds numbers associated to the respective airfoil sections analyzed in the present manuscript are $Re$ = $14.088\times 10^6$ and $Re$ = $14.877\times 10^6$, the characteristic length being the corresponding chord length for each airfoil.}

\keywords{Computational Fluid Dynamics (CFD), Active Flow Control (AFC), Synthetic jet, Boundary layer, Aerodynamic efficiency, Parametric optimization}

\maketitle

\section{Introduction}

The number of articles tackling Active Flow Control (AFC) applied to airfoils is rising sharply \citep{zhu2018simulation, hochhausler2021experimental, khalil2021active, mosca2021multidisciplinary, maldonado2023effect}. The AFC technology basically consists of adding or subtracting momentum to/from the main flow in order to interact with the boundary layer and delay its separation. Its main advantage over passive flow control approaches is that it is not creating drag penalty in off-design conditions. 

The different AFC techniques were divided by \citet{Cattafesta_2011} into three categories. 1) Moving body actuators \citep{Wang_2019}, 2) Plasma actuators \citep{Cho_2011,Foshat_2020, Benard_2014,Benard_2016} and 3) Fluidic actuators (FA). In some particular FA designs, the origin of the self-sustained oscillations was recently unveiled in \citep{bergada2021fluidic,Baghaei_2020, Baghaei_2019}. 
From the different FA, Zero Net Mass Flow Actuators (ZNMFA) also called Synthetic Jet Actuators (SJA), have the advantage of not requiring external fluid supply, being as well particularly effective in controlling the separation of the boundary layer \citep{glezer2002synthetic,rumsey2004summary,wygnanski2004variables,findanis2008interaction}. 
The use of pulsating flow has the advantage of coupling with the boundary layer natural instabilities and so being more energetically efficient \cite{de2015comparison, traficante2016flow, zhang2019comparison}. In fact, and in order to minimize the energy required in AFC implementations it is needed to somehow optimize the different parameters associated to each particular AFC application, nowadays the most common methodology employed to tune the AFC associated parameters is a parametric optimization although the use of optimizers is increasingly accepted \citep{Tousi2021, tadjfar2020optimization, tousi2022large}. 

Active Flow Control implementation in Vertical (VAWT) and Horizontal Axis Wind Turbines (HAWT) is recently gaining momentum \cite{stalnov2010evaluation,maldonado2010active,yen2013enhancing,taylor2015load,velasco2017numerical,zhu2018simulation,moshfeghi2017numerical, maldonado2019increasing,wang2022effect,maldonado2023effect}, although so far, the bases to optimize the five AFC associated parameters along the blade are not being established. Among the different devices employed in AFC applications on (WT), synthetic Jets (SJ) appear to be particularly effective, then it combines the use of periodic forcing (blowing/suction) with a sufficiently large momentum coefficient $C_\mu$, which is defined as $C_{\mu}={(h \rho_{jet}({U}_{j}^{2})\sin\theta})/{(C\rho_{\infty}(U_{\infty}^{2}))}$, where $h$ is the jet width, $\rho_{jet}$ and $\rho_{\infty}$ are the jet and far field densities, respectively, ${U}_{j}$ is the maximum jet velocity, $C$ the airfoil chord, $U_\infty$ the free-stream velocity and $\theta$ is the jet inclination angle with respect to the adjacent surface, combination which appears to be particularly effective in delaying the boundary layer separation. The non-dimensional frequency associated to periodic forcing is defined as, $F^{+}=f{C}/{U_{\infty}}$, being $f$ the boundary layer natural dimensional frequency.   

A thorough experimental work employing SJ on wind turbines blade sections was undertaken by \cite{stalnov2010evaluation}, where it was established that AFC-SJ was capable of delaying airfoils stall and proved that a significant reduction of wind turbine start-up velocity was possible. In \cite{maldonado2010active,maldonado2009active}, SJ were employed to control the fluid flow vibration induced on turbine blades, they observed that the the boundary layer reattachment obtained when applying AFC was having associated a lower amplitude lift and drag dynamic coefficients and with it smaller structural vibrations.      
The use of SJ to enhance the performance of VAWT was experimentally addressed in \cite{yen2013enhancing}, they obtained a considerable reduction of the noise generated while increasing the power generated by the turbine as well as its safety. Similar experimental studies using SJ but on a VAWT S809 and a HAWT S809 airfoils were performed by \cite{taylor2015load} and \cite{maldonado2019increasing}, respectively. The former study demonstrated that lift, drag and pitching moment hysteresis could be significantly reduced. In the later research they observed that a 10.6$\%$ power input reduction to drive the rotor was achieved when a total of 20 SJA along each blade were employed.  

Due to the high cost associated to experimental studies and the increasing trustability of the computational simulations, the nowadays tendency is to simulate the performance of any device. In this direction is going the 2D-URANS numerical simulations on VAWT performed by \cite{velasco2017numerical,zhu2018simulation}. In both cases a straight-bladed WT was investigated although the second one was of Darrieus  type.
Three different SJ locations were investigated in the later research while in the former not just the location but also the number of holes of the SJ were considered. A net power increase was obtained in both investigations, a power coefficient increase of 15.2$\%$ was obtained in \cite{zhu2018simulation}.
A numerical investigation on the S809-HAWT airfoil aerodynamic enhancement using SJ was undertaken by \cite{moshfeghi2017numerical}. Lift increase was observed for a wide range of injected flow angles. As previously performed by \cite{zhu2018simulation}, \citet{wang2022effect} placed the SJA at the airfoil trailing edge and numerically studied the aerodynamic characteristics improvement on a straight-bladed VAWT. They evaluated the performance of several configurations of Dual synthetic Jets Actuators DSJA to improve the power capacity of the WT and observed that the power coefficient could be increased by 58.87$\%$ when the Juxtaposition-type DSJA was used.      
One of the latest research on AFC applications to HAWT with S809 airfoils is the one performed experimentally by \citet{maldonado2023effect}. Two rotors with blade aspect ratios of 7.79 and 9.74 and having 20 SJA distributed along the span of each blade were analyzed at four different turning speeds and three pitch angles. The rotor with a higher aspect ratio was found to be more aerodynamic efficient and thanks to the use of AFC, unsteady bending and unsteady torsion could be reduced by 36.5$\%$ and 22.9$\%$ respectively.  

In all the research just presented on WT, the five AFC parameters, SJ position, width, injection angle, frequency and momentum coefficient, were chosen without performing a previous optimization, which appears to be the necessary task if minimum energy for the actuation is to be used. In fact, when aiming to optimize the different AFC parameters the following considerations gathered from previous investigations are needed.

When experimentally studding the flow control on NACA-0025 airfoil at $Re$ = $10^{5}$ and AoA = $5^\circ$, \citet{goodfellow2013momentum} observed that the momentum coefficient was the primary control parameter. When analyzing the effect of jet position in controlling the boundary layer separation
\citep{feero2017influence}, observed that the maximum effectiveness was obtained when placing the jet groove close to the boundary layer separation point, either downstream or upstream. The same observations were made by {\citet{amitay2001aerodynamic} and \citet{amitay2002role} which observed that locating the actuator near to the boundary layer separation point, a lower momentum coefficient was needed to reattach the separated flow. Furthermore, \citet{feero2015flow} observed that the required $C_\mu$ to reattach the boundary layer was the lowest for excitation frequencies in the range of the vortex shedding frequency.
The same observation was made by 
\citet{tuck2008separation}, then they reported optimal actuation frequencies of $F^{+}$ = 0.7 and $1.3$, the highest one being most effective in combination with an optimum momentum coefficient of $C_{\mu}$ = 0.0123. \citet{kitsios2011coherent} conducted a LES study on a NACA-0015 airfoil at $Re$ = $8.96\times10^{5}$ and noticed that the optimal pulsating frequency coincided with the baseline shedding frequency ($f_{wake}$). This was also experimentally confirmed by \citet{buchmann2013influence} and corroborated using (3D-DNS) by \citet{zhang2015direct}.  
Numerical studies on the NACA-23012 airfoil at $Re$ = $2.19\times10^{6}$ were performed by \cite{kim2009separation, monir2014tangential} and concluded that maximum lift was obtained when $F^{+}$ = 1, the jet was placed nearby the boundary layer separation point, tangential injection/suction was observed to be particularly effective.

The work performed in the present manuscript considers and adapts the previous knowledge on AFC and applies it to set a methodology which shall optimize the AFC SJA parameters to all WT airfoil sections. Via considering the methodology here outlined the energy required to reattach the boundary layer in all WT sections shall be minimized therefore maximizing the energy increase obtained by any WT.

The remainder of the paper is structured as follows. The problem formulation, numerical methods and mesh independence study are presented in section \ref{sec:num_mthd}. Section \ref{dtudefinition} serve to define the WT main characteristics and the airfoils selected to be simulated via CFD. The baseline case study of the different sections, the complete parametric analysis of the chosen airfoil and the energy assessment, are to be found in section \ref{results}. The research performed is summarized in section \ref{sec:Conclusions}.

\section{Numerical Method} \label{sec:num_mthd}

\subsection{Governing equations and turbulence model}

The Navier-Stokes (NS) equations under incompressible flow conditions take the form:

\begin{equation}
\frac{\partial {u}_{i}}{\partial x_{i}}=0
\end{equation}

\begin{equation}
\frac{\partial {u}_{i}}{\partial t}+\frac{\partial {u}_{i} {u}_{j}}{\partial x_{j}}=-\frac{1}{\rho} \frac{\partial {p}}{\partial x_{i}}+\nu \frac{\partial^{2} {u}_{i}}{\partial x_{j} \partial x_{j}}
\end{equation}

As Reynolds number increases, the mesh and the associated time step NS equations require drastically decreases in order to reach the Kolmogorov length and time scales, but reaching such scales involve extremely large computational times, such simulations are called Direct Numerical Simulations (DNS). Due to its drawbacks, nowadays DNS is just used in research purposes and for relatively small Reynolds numbers.    
For highly turbulent flow and to obtain reasonable accurate results, turbulence models are still needed. The use of Large Eddy simulation (LES) as turbulence model, is a very good option in 3D flows, yet it still requires very fine meshes and large computational times, therefore LES simulations require supercomputers to shorten the computational time. As a result, simulations at large Reynolds numbers still need to be performed using Reynolds Averaged Navier Stokes (RANS) or URANS unsteady-RANS turbulence models. Their precision is not as accurate as LES or DNS, but the computational time needed shortens drastically. 

Under incompressible flow conditions, the only variables associated to the NS equations are the pressure and the three velocity components, which are generically called $\phi$. In order to be able to apply URANS models, each variable from the NS equations need to be substituted by its average $\bar{\phi}$ and a fluctuation $\phi^{\prime}$term.



\begin{equation}
    \phi=\bar{\phi}+\phi^{\prime}
\end{equation}


After the substitution, the resulting NS equations (considering just two dimensions) take the form:

\begin{equation}
    \frac{\partial \rho}{\partial t}+\rho \left ( \frac{\partial  \bar{u}_x}{\partial x}+\frac{\partial \bar{u}_y}{\partial y}\right )=0
    \label{eq:contRANS}
\end{equation}

\begin{equation}
     \frac{\partial \bar{u}_x}{\partial t}+\bar{u}_x\frac{\partial \bar{u}_x}{\partial x}+\bar{u}_y\frac{\partial \bar{u}_x}{\partial y} =-\frac{1}{\rho} \frac{\partial p}{\partial x}+ g_x+ \frac{1}{\rho}\frac{\partial }{\partial x}\left ( 2 \mu\frac{\partial \bar{u}_x}{\partial x}-\rho (\bar{u}'_x)^2 \right )+\frac{1}{\rho} \frac{\partial }{\partial y}\left ( \mu \left ( \frac{\partial \bar{u}_x}{\partial y}+\frac{\partial \bar{u}_y}{\partial x} \right ) -\rho\overline{u'_xu'_y} \right )
    \label{eq:contRANS-2}
\end{equation}

\begin{equation}
     \frac{\partial \bar{u}_y}{\partial t}+\bar{u}_x\frac{\partial \bar{u}_y}{\partial x}+\bar{u}_y\frac{\partial \bar{u}_y}{\partial y} =-\frac{1}{\rho} \frac{\partial p}{\partial y}+ g_y+ \frac{1}{\rho}\frac{\partial }{\partial x}\left ( \mu \left ( \frac{\partial \bar{u}_y}{\partial x}+\frac{\partial \bar{u}_x}{\partial y} \right ) -\rho\overline{u'_yu'_x} \right )+\frac{1}{\rho} \frac{\partial }{\partial y}\left ( 2 \mu\frac{\partial \bar{u}_y}{\partial y}-\rho (\bar{u}'_y)^2 \right )
    \label{eq:contRANS-3}
\end{equation}
At this point it is worth introducing the apparent Reynolds stress tensor ($\tau_{app}$), which comprises a symmetric matrix encompassing the averaged fluctuations in the $x$ and $y$ directions. It is important to observe that in the case of two-dimensional flows, this matrix explicitly consists of four terms, whereas in three dimensions, it expands to a $3\times 3$ symmetric matrix.

All RANS turbulence models are based on solving the Navier-Stokes equations by incorporating the concept of turbulence viscosity ($\mu_t$), which can be mathematically implemented into the momentum equations described above using the subsequent definition (called the Boussinesq hypothesis):

\begin{equation}
    \mu_t=\frac{-\rho \overline{u'_xu'_x}}{2\dfrac{\partial \bar{u}_x}{\partial x}}=\frac{-\rho \overline{u'_yu'_y}}{2\dfrac{\partial \bar{u}_y}{\partial y}}=\frac{-\rho \overline{u'_xu'_y}}{\dfrac{\partial \bar{u}_x}{\partial y}+\dfrac{\partial \bar{u}_y}{\partial x}}
    \label{eq:mu-t}
\end{equation}

The X and Y momentum terms of the NS equations can therefore be expressed as: 

\begin{equation}
    \dfrac{\partial \bar{u}_x}{\partial t}+ \bar{u}_x \dfrac{\partial \bar{u}_x}{\partial x}+ \bar{u}_y\dfrac{\partial \bar{u}_x}{\partial y}= g_x-\frac{1}{\rho} \dfrac{\partial p}{\partial x}+\frac{1}{\rho}\dfrac{\partial }{\partial x}\left ( 2(\mu+\mu_t)\dfrac{\partial \bar{u}_x}{\partial x} \right )
    +\frac{1}{\rho}\dfrac{\partial }{\partial y}\left ( (\mu+\mu_t)\left ( \dfrac{\partial \bar{u}_x}{\partial y}+\dfrac{\partial \bar{u}_y}{\partial x} \right ) \right )
\label{eq:momxRANS}
\end{equation}

\begin{equation}
    \dfrac{\partial \bar{u}_y}{\partial t}+\bar{u}_x \dfrac{\partial \bar{u}_y}{\partial x}+\bar{u}_y\dfrac{\partial \bar{u}_y}{\partial y}=g_y-\frac{1}{\rho}\dfrac{\partial p}{\partial y}+\frac{1}{\rho}\dfrac{\partial }{\partial x}\left ( (\mu+\mu_t)\left ( \dfrac{\partial \bar{u}_x}{\partial y}+\dfrac{\partial \bar{u}_y}{\partial x} \right ) \right )
    +\frac{1}{\rho}\dfrac{\partial }{\partial y}\left ( 2(\mu+\mu_t)\dfrac{\partial \bar{u}_y}{\partial y} \right )
\label{eq:momyRANS}
\end{equation}

We have chosen to use the $k-\omega$ SST turbulence model which rely on using the $k-\omega$ model near the wall, the $k-\epsilon$ model far away from the object and a blending function between these two. Mathematically $\mu_t$ is calculated as:

\begin{equation}
    \mu_t = \frac{\rho k}{\omega} \:\:\:\:\:\: \longrightarrow \:\:\:\:\:\: \begin{cases}
\rho: \text{density} \\
k: \text{turbulent kinetic energy} \\
\omega: \text{turbulent kinetic energy specific dissipation rate} 
\end{cases}
\label{eq:mu-t-2}
\end{equation}

As stated, this model is a two transport equation model in order to solve $k$ and $\omega$. The equations used for each parameter are:
\begin{equation}
    \frac{\partial k}{\partial t}+u_j\frac{\partial k}{\partial x_j}=P_k-\beta^*k\omega +\frac{\partial }{\partial x_j}\left [ \left ( \nu +\sigma_k\nu_T \right ) \frac{\partial k}{\partial x_j} \right ]
\end{equation}

\begin{equation}
    \frac{\partial \omega}{\partial t}+u_j\frac{\partial \omega}{\partial x_j}=\alpha S-\beta\omega^2 +\frac{\partial }{\partial x_j}\left [ \left ( \nu +\sigma_\omega \nu_T \right ) \frac{\partial \omega}{\partial x_j} \right ]+2\left ( 1-F_1 \right )\sigma_{\omega 2}\frac{1}{\omega}\frac{\partial k}{\partial x_i}\frac{\partial \omega}{\partial x_i}
\end{equation}

Where $F_1$ and $F_2$ are the blending functions and numerically depend on the distance of the cell to the wall. The first blending function takes the value of 0 far away from the wall, 1 in the cells close to the wall and values between 0 and 1 in the transition region. The second blending function $F_2$ depends also on the perpendicular distance from the wall ($d$) and according to \citep{Menter}, ``since the modification to the eddy-viscosity has its largest impact in the wake region of the boundary layer, it is imperative that $F_2$ extends further out into the boundary-layer than $F_1$''. Mathematically is expressed as:
\begin{equation}
    F_2 = \tanh \left ( \arg^2_2 \right )
\end{equation}
\begin{equation}
    \arg^2_2 = \max\left ( \frac{2k}{\beta^*\omega d};\: \frac{500 \nu}{\omega d^2} \right )
\end{equation}

The constants of the $\arg$ function are adjusted manually, and for the present case the values of the used turbulence model taken are: $\alpha_1=0.556$, $\alpha_2=0.44$, $\beta^*=0.09$, $\beta_1=0.075$, $\beta_2=0.0828$ $\sigma_{k1}=0.85$, $\sigma_{k2}=1$, $\sigma_{\omega1}=0.5$, $\sigma_{\omega2}=0.856$. For a more detailed information of the $k-\omega\:SST$ turbulence model proposed, the reader should address to \citep{Menter}.

\subsection{Non-dimensional parameters}

\textbf{Reynolds Number}

The Reynolds number is defined as the ratio of inertial to viscous forces. Mathematically it is expressed as:
\begin{equation}
    Re=\frac{\rho u_{rel} C}{\mu}
    \label{eq:Reynolds}
\end{equation}

Where $u_{rel}$ is the relative velocity, $C$ the chord length and $\mu$ the fluid absolute viscosity.

\textbf{Aerodynamic Coefficients}

The most common aerodynamic coefficients are the lift $C_l$ and drag $C_d$ ones which are mathematically expressed as follows:
\begin{equation}
    C_l=\frac{L}{\dfrac{1}{2}\rho u_{rel}^2C} \hspace{25mm} C_d=\frac{D}{\dfrac{1}{2}\rho u_{rel}^2C} 
    \label{eq:AeroCoeffs}
\end{equation}
L and D represent the lift and drag forces, respectively.

The aerodynamic efficiency is the ratio between the lift and drag coefficients:
\begin{equation}
    \eta = \frac{C_l}{C_d}
    \label{eq:E}
\end{equation}

\textbf{Pressure Coefficient}

It characterizes the relative pressure field ($p-p_{\infty}$) along the body.
 
It is mathematically expressed as:

\begin{equation}
    C_p = \frac{p-p_{\infty}}{\frac{1}{2}\rho u_{rel}^2}
    \label{eq:Cp}
\end{equation}
Where ($(1/2)\rho u_{rel}^2$) is defined as the dynamic pressure.

\textbf{Friction Coefficient}

The friction coefficient, also known as skin friction coefficient or tangential friction coefficient, evaluates the frictional force per unit area $\tau_w$ acting on a surface in contact with the fluid.
\begin{equation}
    C_{f} = \dfrac{\tau_w}{\frac{1}{2}\rho u_{rel}^2}
    \label{eq:Cf}
\end{equation}

\textbf{Momentum Coefficient}

It represents the momentum associated to the jet divided by the incoming fluid momentum. According to reference \cite{goodfellow2013momentum} $C_{\mu}$ is the primary AFC parameter. Mathematically it is expressed as:
\begin{equation}
    C_\mu = \frac{h\left ( \rho u_{max}^2  \right )\sin \theta_{\text{jet}}}{C\left ( \rho u_{rel}^2  \right )}
    \label{eq:Cmu}
\end{equation}

Where ($h$) is the jet width, ($u_{max}$) is the maximum jet velocity and ($\theta_{\text{jet}}$) is the jet inclination angle with respect to the airfoil surface.

\textbf{Forcing Frequency}

The forcing frequency given as a non dimensional number is defined as follows:
\begin{equation}
    F^+= \frac{f}{f_0}
    \label{eq:forcfreq}
\end{equation}
Where $f$ is an arbitrary frequency and $f_0$ is the vortex shedding frequency (the frequency at which vortices are shed from the airfoil).

\textbf{Courant-Friedrichs-Levy Number}

The Courant number, often denoted as $\mathrm{CFL}$ refers to the ability of the system to capture the field of interest. Mathematically it is defined as the fluid velocity measured at any given mesh cell multiplied by the simulation time step and divided by the mesh cell length scale ($\Delta x$). 
\begin{equation}
    CFL = \frac{{u} \Delta t}{\Delta x}
    \label{eq:CFL}
\end{equation}
For time-dependent simulations, the Courant number is an important parameter in order to ensure stability of the system, and it is recommended to keep it below 1.

\textbf{Dimensionless wall distance }

The dimensionless wall distance $y^+$ represents the distance from the wall to the first mesh cell, $y$, normalized by a characteristic velocity $u_{\tau}$ and the relative viscosity of the fluid $\nu$. Mathematically it is expressed as:
\begin{equation}
    y^+=\frac{u_\tau y}{\nu}
    \label{eq:yplus}
\end{equation}

Where $u_\tau$ is the friction velocity, defined as:
\begin{equation}
    u_\tau = \sqrt{\frac{\tau_w}{\rho}}
\label{eq:utau}
\end{equation}

\subsection{Numerical domain and boundary conditions}

For computational domains relatively small in respect to the bluff-body characteristic length, the flow around the bluff-body is affected by the boundary conditions defined at the boundaries. Conversely, if the domain is chosen to be too large, there may be excessive mesh cells, resulting in longer computational times and poor simulation efficiency. Therefore, it is crucial to choose an appropriate domain size that balances the need for accurate results with computational efficiency.

Based on the previous research undertaken by \cite{couto2022aerodynamic,Tousi2021,tousi2022large} the computational domain employed in the present research has the following dimensions, $10C$ from the domain inlet to the airfoil leading edge, $15C$ from the trailing edge to the domain outlet, and $10C$ from the body surface to both the upper and lower limits of the domain, $C$ being the airfoil chord, see Fig. \ref{fig:domaindiagram}.

\begin{figure}[]
    \centering
    \includegraphics[width=0.8\textwidth]{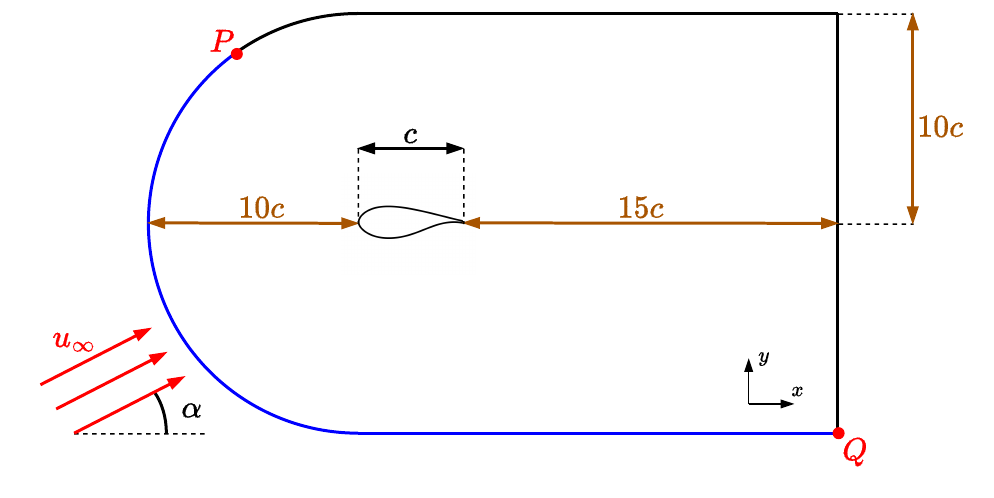}
    \caption{Schematic view of the meshing domain within its general dimensions.}
    \label{fig:domaindiagram}
\end{figure}

Two points $P$ and $Q$ delimiting the inlet and outlet boundaries have been implemented in Fig. \ref{fig:domaindiagram}. The outlet boundary extends from $P$ to $Q$ on clockwise direction, and from $P$ to $Q$ on counterclockwise direction is to be found the inlet boundary.
The curvilinear shape of the domain at the inlet, has been selected to facilitate the simulation of various angles of attack without altering the mesh geometry. It should be noted that the airfoil has been horizontally positioned, and by adjusting the inlet velocity components, any required angle of attack can be obtained. This approach not only allows flexibility in the simulation but also ensures that the mesh remains consistent throughout the analysis.

The present research focuses on simulating external aerodynamics with transient turbulent flow (aerodynamic forces change over time) and at large Reynolds numbers of the order of $\mathcal{O}(10^7)$. 
Additionally, the angle of attack is relatively high for the two baseline case sections studied, which particularly for the section placed nearest the blade root (section 54 of the DTU-10MW-HAWT), it is expected to produce the separation of the boundary layer at some point along the chord, generating the corresponding vortical structures and vortex shedding. Moreover, since the relative velocity in both sections studied does not exceed the incompressible flow limit, which is commonly accepted as $M<0.3$, the flow will be considered as incompressible across the entire study.
The flow solver contained in the OpenFOAM package, which fulfills the needs for the present research is \textit{PisoFoam} (Pressure implicit with splitting operator). This solver is particularly well-suited for simulating unsteady turbulent flow, and is widely used in aerodynamic simulations due to its accuracy and efficiency.

\textcolor{black}{Section 54 with a wind speed of $10m/s$ was chosen to perform the mesh independence study. Notice that the Reynolds number is almost the same in both sections and the angle of attach is much larger in this particular one}.
The main data relative to the two sections studied, section 54 and 81 is presented in Table \ref{tab:BCSparam70}, which includes the airfoil type, the radius versus the turbine main axis, the section number, the chord length, the Reynolds number, the angle of attack, and the relative velocity, which is the vectorial magnitude obtained from the combination of the wind and turning speeds and when considering the axial (a) and angular (a') induction factors. Figure \ref{sections} introduces the approximate location of the two sections under study as well as their respective profile.

\begin{table}[]
\centering
\renewcommand{\arraystretch}{1.5}
\setlength{\tabcolsep}{16pt}
\begin{tabular}{ccc}
\hline
\textbf{Airfoil Type}                     & FFA-W3-301 & FFA-W3-301   \\
\textbf{Radius ($z$)}                     & $31.42\:m$  & $53.43\:m$  \\
\textbf{Section Number ($N$)}             & $54$   & $81$          \\
\textbf{Chord length ($c$)}               & $6.06\:m$ & $4.44\:m$    \\
\textbf{Reynolds number ($Re$)}          & $14.08\times10^6$  & $14.87\times10^6$ \\
\textbf{Angle of Attack ($\alpha$)}       & $17.79\:deg.$ & $10.01\:deg.$ \\
\textbf{Freestream velocity ($u_{rel}$)} & $34.40\:m/s$  & $49.61\:m/s$ \\ \hline
\end{tabular}
\caption{main characteristics of the two sections studied. Section 54 and section 81.}
\label{tab:BCSparam70}
\end{table}

\begin{figure}[h]
    \centering
    \includegraphics[width=0.8\textwidth]{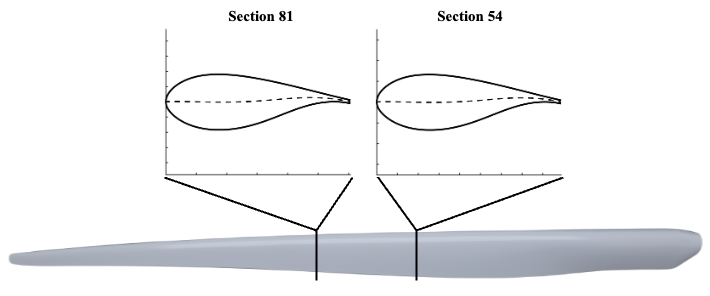}
    \caption{3D view of the DTU10MW blade with the two sections studied.}
    \label{sections}
\end{figure}

The different boundary conditions employed to simulate section 54 and regardless of the mesh density considered are outlined in Table \ref{table2}. It's worth noting that these specific values are based on the $k-\omega\:SST$ turbulence model, aligned with the corresponding chord length and a turbulence intensity at the inlet of $0.1\:\%$ and according to \citep{Menter}.
In this paper, all simulations were performed using air at sea level based on the International Standard Atmosphere (ISA). The two relevant air parameters, density, and kinematic viscosity are summarized in Table \ref{tab:airparam}.

\begin{table}[]
\centering
\renewcommand{\arraystretch}{1.25}
\setlength{\tabcolsep}{4pt}
\begin{tabular}{p{1.4cm} p{1.7cm} p{2.9cm} p{3.4cm} p{1.8cm} p{1.8cm} }
\hline
 & \multicolumn{5}{c}{Simulation Parameters} \\ 
  \hline
& $k (m^2/s^2)$ & $omega (s^{-1})$  & $nut (m^2/s)$ & $P (Pa)$ & $u_{rel} (m/s)$ \\
\hline
 inlet  &  $1.78 \times 10^{-3}$  &  $6.95 \times 10^{-3}$ & $2.25 \times 10^{-1}$ &  zeroGradient & $u_x=32.75$; $u_y=10.51$  \\
   \hline
 outlet    &  zeroGradient & zeroGradient & zeroGradient & 0 & zeroGradient   \\
   \hline
 front $\&$ back & empty  & empty & empty & empty & empty \\
  \hline
wall  & $10^{-20}$ & ohmegaWallFunction value $10^{-5}$ & nutLowReyWallFunction value 0 & zeroGradient & 0  \\
  \hline
\end{tabular}
\caption{Summary of the boundary conditions for the baseline case. Section 54 at $u_\infty=10\:m/s$.}
\label{table2}
\end{table}

\begin{table}[]
\centering
\renewcommand{\arraystretch}{1.5}
\setlength{\tabcolsep}{20pt}
\begin{tabular}{cc}
\hline
\textbf{Density ($\rho$)}            & $1.225 \: kg/m^3$ \\
\textbf{Kinematic viscosity ($\nu$)} & $1.48\cdot 10^{-5}\: m^2/s$    \\ \hline
\end{tabular}
\caption{Density and kinematic viscosity values at sea level and according to ISA conditions.}
\label{tab:airparam}
\end{table}

\subsection{Mesh assessment}


The blade considered in the present research is the DTU-10MW offshore HAWT and consist of six different airfoil types, which seen from the central axis to the blade tip, see Figure \ref{fig:DTUBlade}, are: FFA-W3-600; FFA-W3-480; FFA-W3-360; FFA-W3-301; FFA-W3-241 and NACA0015. Note that DTU implemented a Gurney flap at the trailing edge of the FFA-W3-600 airfoil type, so in terms of meshing, this modifications must be considered. 
\textcolor{black}{The mesh independence study was performed using section 54 which belongs to FFA-W3-301 airfoil type. For a wind speed of $10m/s$ the Reynolds number associated to this section is $Re=14.08\times10^6$ and considering the angle of attack associated, boundary layer separation is expected to occur.}

\begin{figure}[]
    \centering
    \includegraphics[width=1\textwidth]{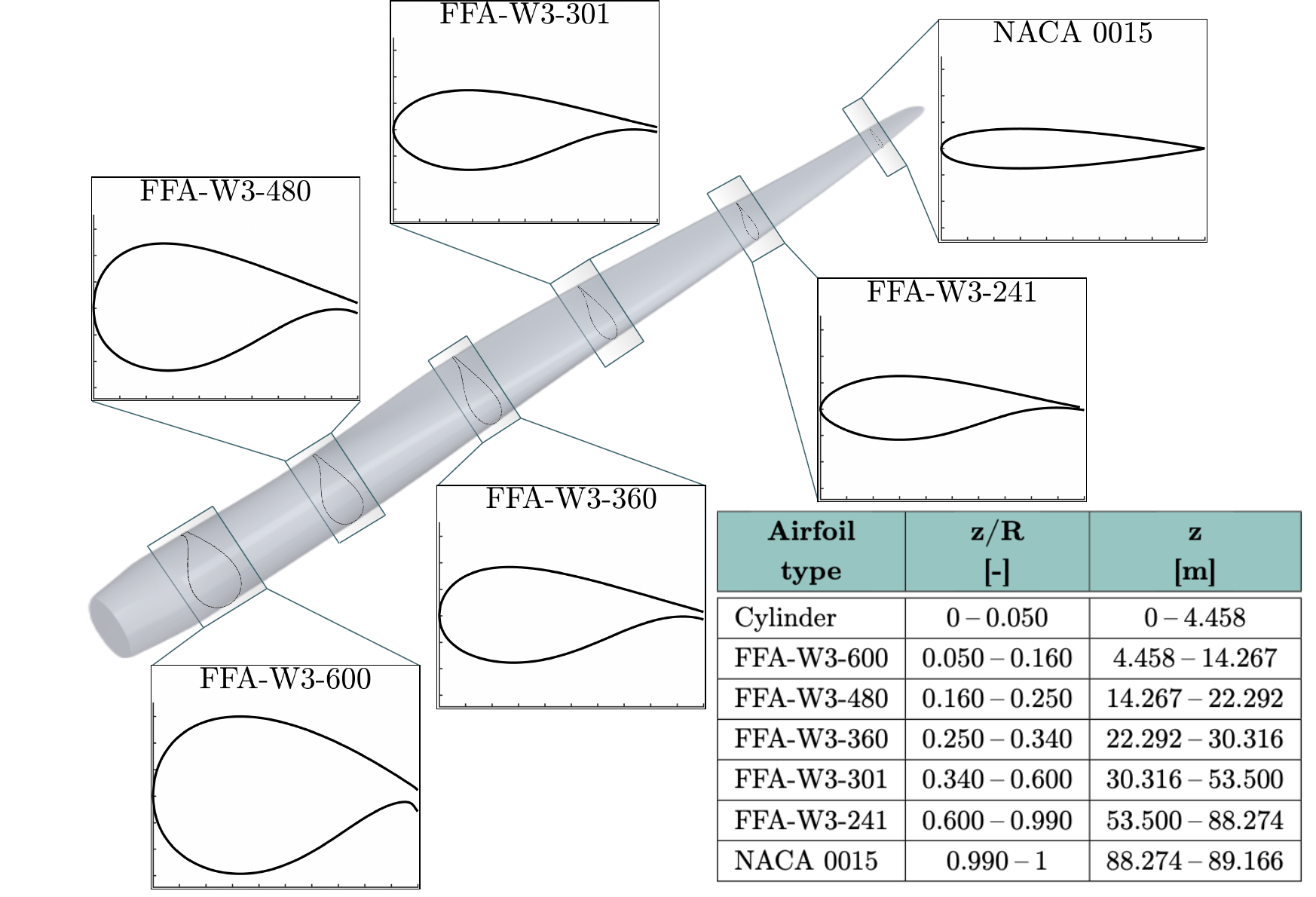}
    \caption{Blade diagram with the different airfoil profiles along the blade span.}
    \label{fig:DTUBlade}
\end{figure}

For the FFA-W3-301 airfoil family a C-type body fitted structured mesh has been chosen for the region close to the airfoil. The C-type mesh comprises a curvilinear line surrounding the solid geometry, which it is closed on the leading edge of the airfoil but open on the trailing edge. The region between the airfoil and the C-line is where the structured mesh is allocated, this region is usually called ``halo''. The halo thickness employed in the present research and measured perpendicular to the airfoil surface is about $3.2\%$ of the chord, this value was kept constant for the two airfoils considered. A general view of the computational domain with the hybrid mesh employed is presented in Figure \ref{subfig:meshgen}. A zoomed view of the structured mesh zones at the airfoil leading and trailing edges is presented in Figure \ref{subfig:meshzoomed}, where the details of the mesh at the trailing edge are observed. A small rectangular structured domain has been employed at the trailing edge region to properly capture the vortical structures due to possible flow detachment.
The remaining portion of the computational domain has been designated with a non-structured mesh, see Figure \ref{subfig:meshgen}. This choice is rooted in the understanding that a certain level of reduced accuracy in regions distant from the airfoil surface is acceptable. Furthermore, opting for a hybrid-mesh enables a more favorable balance between computational efficiency and accuracy of the results. Notwithstanding this, a heightened cell density has been strategically implemented in the vicinity of the airfoil to attain the desired level of precision.

The implementation of the AFC technology using synthetic jets (SJ) requires the modification of the mesh in the location where the SJ is to be implemented. Figure \ref{fig:AFCmesh}b presents a zoom view of the mesh around the SJ, and Figure \ref{fig:AFCmesh}a, introduces the main SJ associated parameters which will be tuned via the corresponding parametric optimization.

\begin{figure}
      \centering
	   \begin{subfigure}{0.56\linewidth}
		\includegraphics[width=1\linewidth]{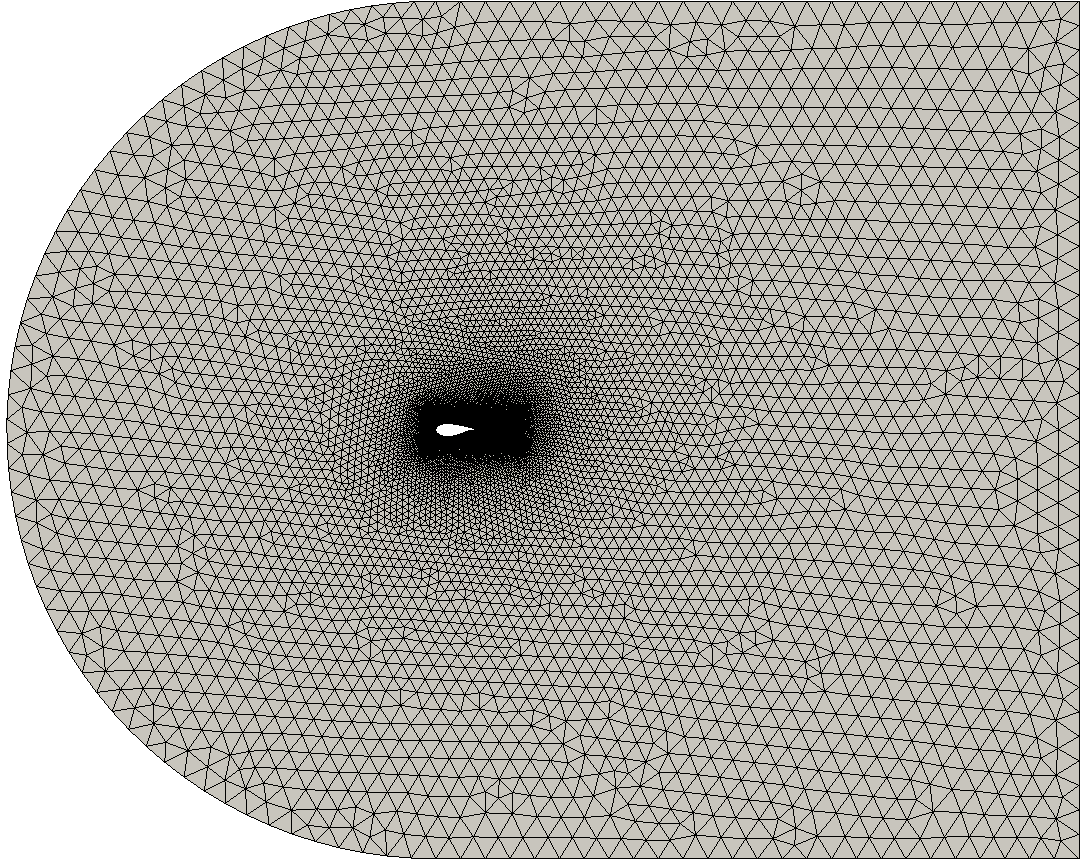}
		\caption{Overall view}
		\label{subfig:meshgen}
	   \end{subfigure}
	     \begin{subfigure}{0.433\linewidth}
	    \includegraphics[width=0.9\linewidth]{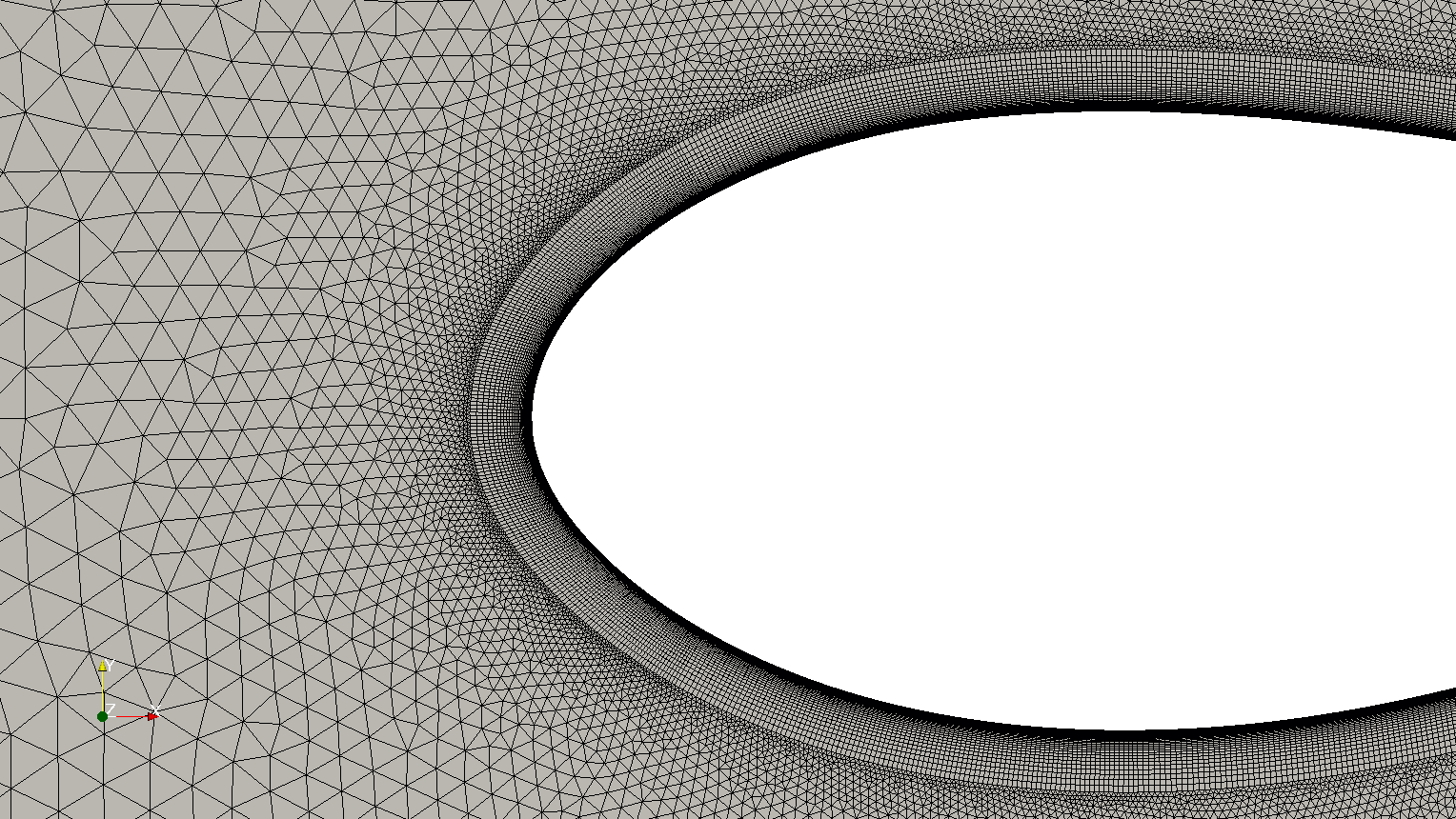}
     \par \smallskip 
	    \includegraphics[width=0.9\linewidth]{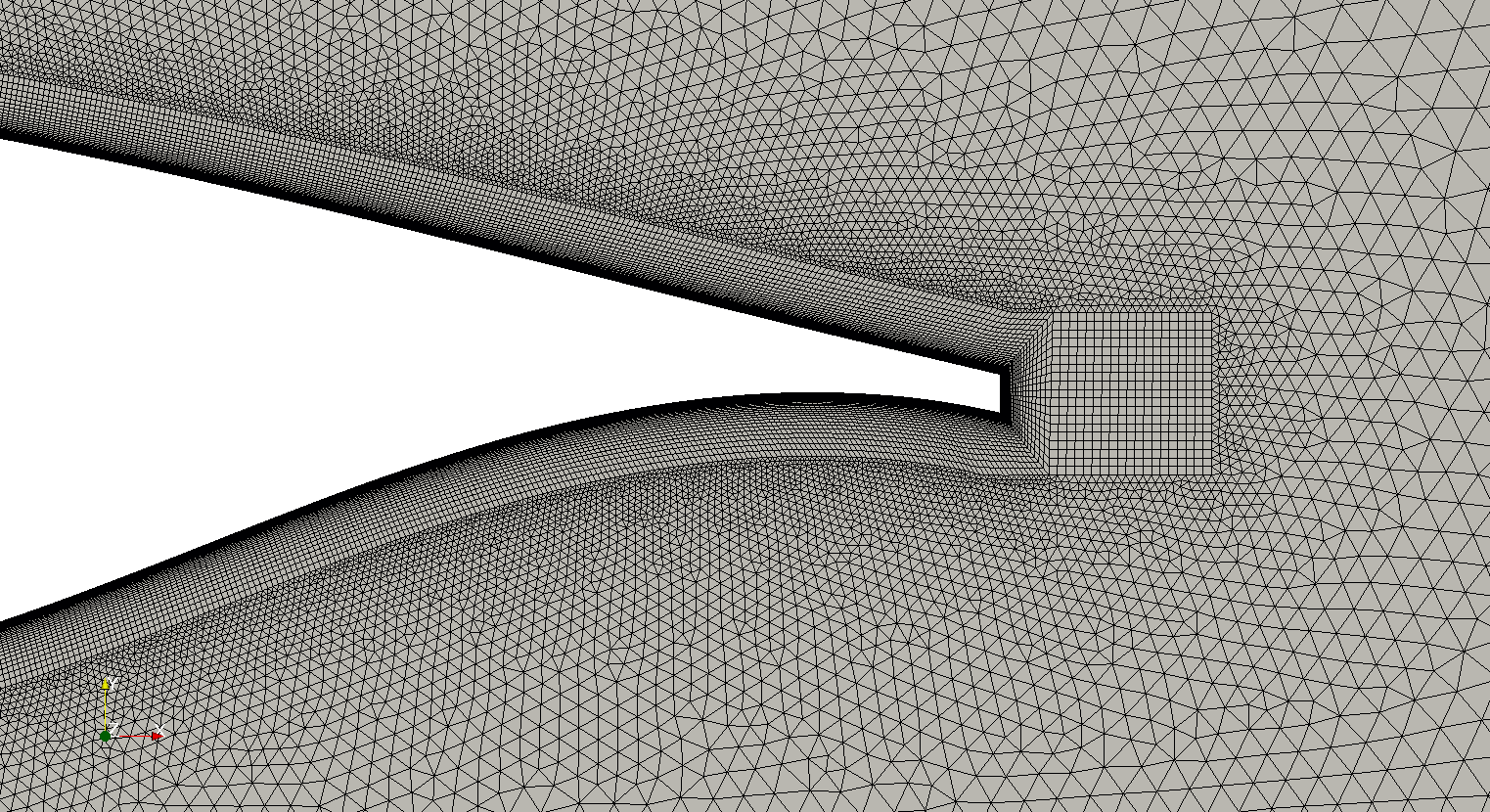}
	    \caption{Zoomed views}
	     \label{subfig:meshzoomed}
	\end{subfigure}
	\caption{Full view of the mesh in the computational domain (a), zoomed view of the structured mesh in the leading (b top) and trailing (b bottom), edge for FFA-W3-301 airfoil type.}
	\label{fig:subfigures4}
\end{figure}

\begin{figure}
    \centering
	   \begin{subfigure}{0.45\linewidth}
		\includegraphics[width=1\linewidth]{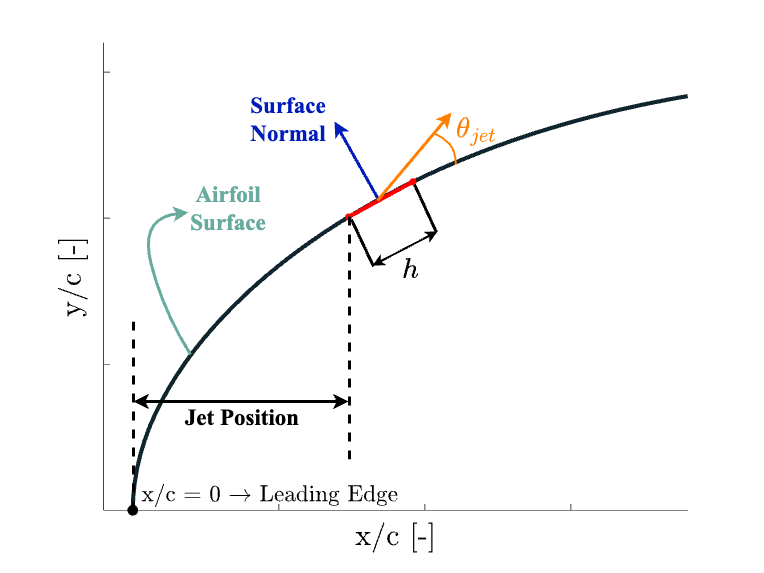}
		\caption{}
            \label{subfig:MeshAFCdet}
	   \end{subfigure}
	     \begin{subfigure}{0.45\linewidth}
	    \includegraphics[width=1\linewidth]{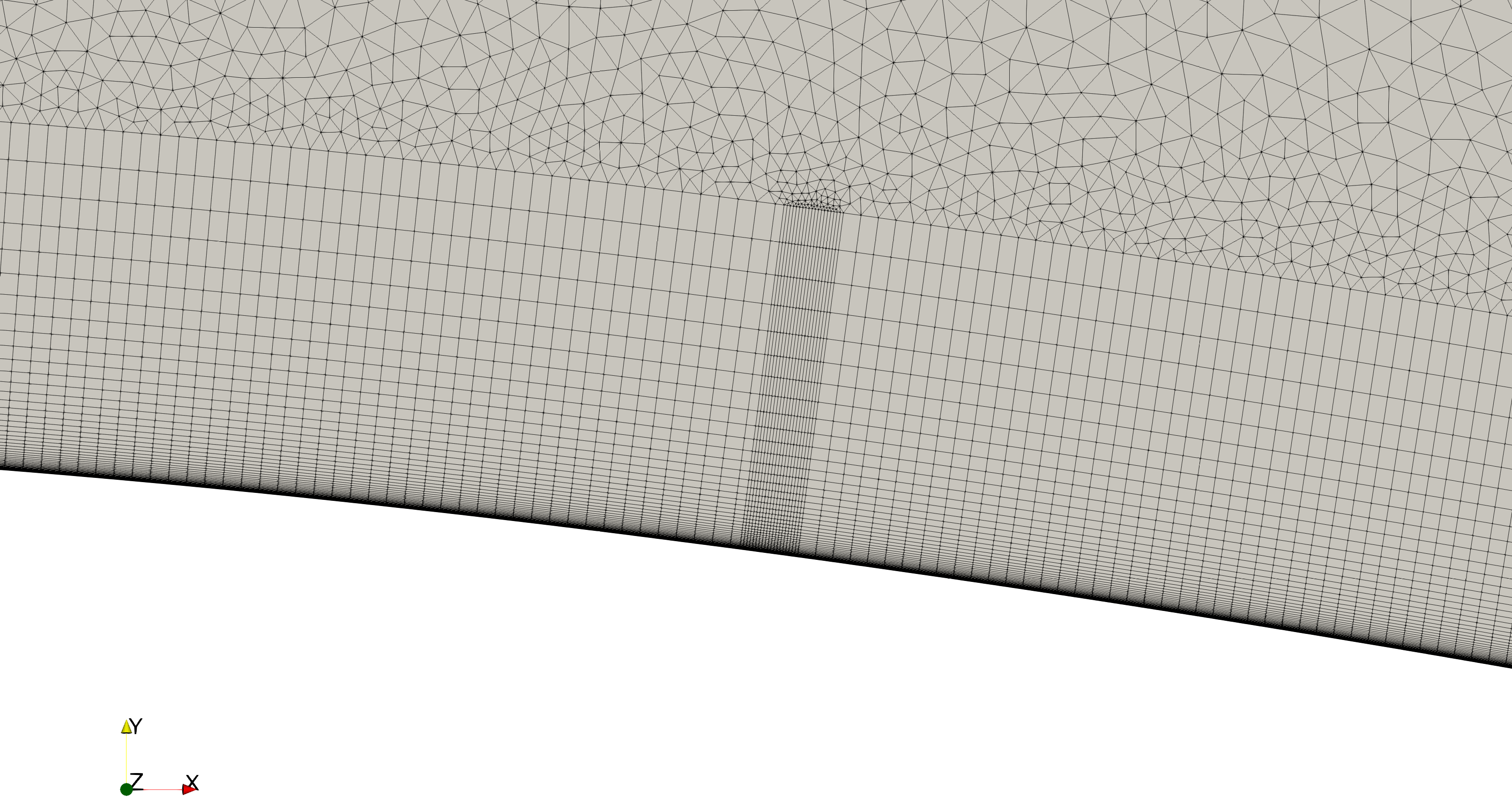}
	    \caption{}
            \label{subfig:MeshAFCgen}
	\end{subfigure}
    \caption{Detailed view of the AFC mesh implementation in Section 54.}
    \label{fig:AFCmesh}
\end{figure}

To ensure a maximum accuracy in the simulated results, a thorough investigation of the mesh density is required, particularly in the halo region near the airfoil surface where the boundary layer develops and it is likely to separate. In this region, a drastic velocity and shear stresses evolution in the normal direction to the airfoil surface is to be expected. 
Due to the lack of prior experimental investigations considering the Reynolds number and the airfoil type chosen and noting that the previous numerical investigations performed by reference \citep{zahle2014comprehensive}, presented the time averaged value for $C_l$, a mesh independence study was performed to ensure a dissociation between mesh resolution and the outcomes produced, thus establishing a robust foundation for further analysis.
The assessment for mesh independence involves generating four distinct meshes, all sharing identical overall geometries (depicted in Figure \ref{subfig:meshgen}), while varying the targeted $y^+$ values. For the purpose of result validation, the approach chosen involves examining the time-averaged $C_l$ and $C_d$, as well as the distributions of $C_p$ and $C_{f}$ along the chord for each mesh resolution. This comparative study seeks to identify the $y^+$ value at which substantial result deviations are absent.

The simulations carried encompassed a $28\:s$ duration, wherein the average lift and drag coefficients were derived from a sampling window spanning the concluding $8\:s$ of simulation. The numerical outcomes of the averaged aerodynamic coefficients, coupled with the corresponding relative error denoted as Error$C_{l}$ in percentage and versus the value obtained by \citep{zahle2014comprehensive}, are presented comprehensively in Table \ref{table:4}. The number of cells used for each mesh, the minimum and maximum $y+$ at the airfoil surface, the airfoil efficiency $\eta$ as well as the boundary layer separation point denoted as $x_s/C$, are presented in the same table. Upon careful examination, it becomes evident that as the maximum $y^+$ is reduced, the error in the lift coefficient drastically drops, becoming negligible between the two finest meshes. For instance, the discrepancy stands at $4.45\%$ between the $C_l$ obtained for maximum $y^+=4.29$ and the value obtained by \citep{zahle2014comprehensive}, such discrepancy is reduced to $2.51\%$ when comparing the reference value and the one obtained in meshes C and D. It is relevant to note that meshes C and D generate identical results for $C_l$ and for the boundary layer separation $x_s/C$, and minor discrepancies are observed in the $C_d$ values generated by these two meshes. Considering the notable accuracy achieved by Mesh C and the manageable computational time entailed in its simulation, it has been judiciously chosen as the preferred candidate for subsequent simulations.

\begin{table}[]
    \centering
\begin{tabular}{l*{9}{c}r} 
\hline
Mesh & $N_{cell}$ &Min $Y^+$ & Max $Y^+$ & $x_s/C$ & $C_l$ & $C_l$ Error [$\%$] & $C_d$  & $\eta (C_l/C_d)$   \\
\hline
A &  $100K$ & $0.513$ & $4.29$ & $0.39$ & $1.5312$&  4.45 & $0.11975$ & $12.79$   \\
B & $172K$ & $0.169$ &  $0.98$ & $0.41$ & $1.5408$&  3.85 & $0.11881$ &  $13.04$   \\
C & $191K$ &$0.064$ &  $0.71$ & $0.42$ &  $1.5623$& 2.51 & $0.10625$ &  $14.71$   \\
D &  $215K$ &$0.039$ & $0.49$ & $0.42$ &  $1.5623$& 2.51 & $0.09985$ &  $15.64$   \\
Zahle et al. \cite{zahle2014comprehensive} &    &  &   &   & 1.60256  & - &     &    \\
\hline
\end{tabular}
\caption{Time-averaged aerodynamic coefficients for different meshes at wind speed $= 10[m/s]$.}
\label{table:4}
    \label{tab:my_label}
\end{table}

For a visual and perhaps more precise comparison of the outcomes, an illustrative representation has been generated to showcase the variation in pressure and friction coefficients across the chord length and for the four distinct tested meshes, as exhibited in Figures \ref{subfig:MIT_cp} and \ref{subfig:MIT_cf}.
Examining the pressure distribution over both the upper and lower surfaces of the airfoil reveals a remarkably uniform behavior across all the meshes. Notably, there exists a consistent and gradual pressure recovery on the upper surface. The sole discernible divergence in pressure coefficients among all the meshes manifests near the $1.5\:m$ mark along the chord. Nevertheless, the graph depicting $C_p$ presents a strikingly analogous pattern for all the meshes.
In the realm of the friction coefficient graph, as depicted in Figure \ref{subfig:MIT_cf}, distinctions emerge among the meshes. Notably, the lower $y^+$ meshes, namely Mesh A and B, exhibit notable variations in the boundary layer separation point, $x_s/C$. Conversely, Mesh C closely mirrors the $C_{f}$ trends observed in Mesh D, aligning with the minor errors observed in the averaged aerodynamic coefficients. This alignment renders Mesh C the optimal selection for future simulations.

\begin{figure}
    \centering
    \begin{subfigure}[t]{0.45\textwidth}
        \centering
        \includegraphics[width=\textwidth]{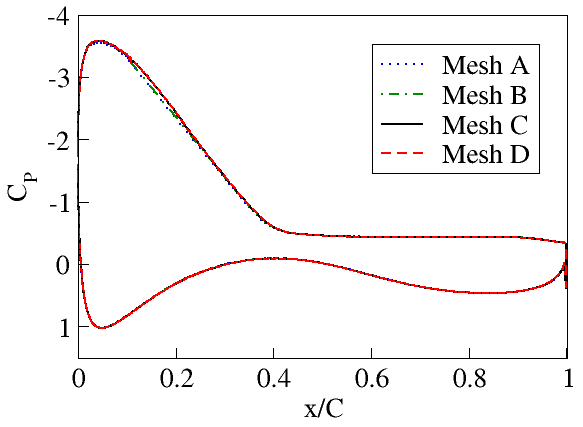}
        \caption{}
        \label{subfig:MIT_cp}
    \end{subfigure}%
    ~ 
    \begin{subfigure}[t]{0.5\textwidth}
        \centering
        \includegraphics[width=\textwidth]{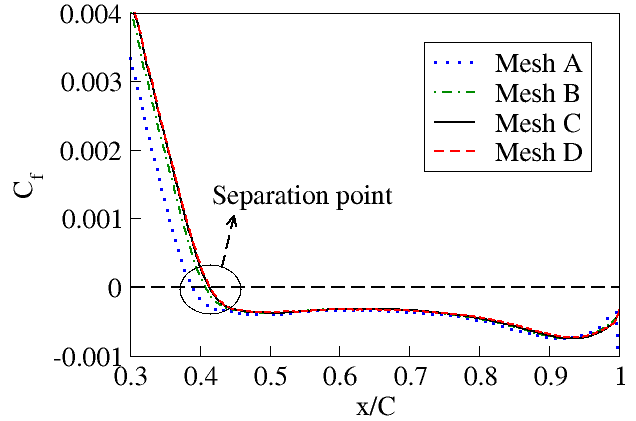}
        \caption{}
        \label{subfig:MIT_cf}
    \end{subfigure}
    \caption{Pressure (a) and friction (b) coefficients comparison for the four meshes evaluated. Section $54$.}
    \label{fig:Cp_comp}
\end{figure}

\section{Main parameters definition of the DTU 10MW RWT} \label{dtudefinition}

In order to increase the accuracy of the 2D-CFD simulations, it is important to consider the flow three dimensional effects over the airfoil section under study. Two induction factors, axial $(a)$ and tangential (angular) $(a')$, which consider some flow three-dimensional effects, are used to more precisely calculate the axial and radial velocities for 2D simulations \cite{treballinductionfactors}. The decrease in streamwise velocity from the freestream conditions to the rotor plane is called the axial velocity deficit, and the fraction by which it is reduced is called the axial induction factor ($a$). The tangential induction factor (a') considers the increase of the tangential speed in the rotor plane. Figure \ref{fig:wind_rotor_velo} introduces the typical velocity and force components acting on a generic HAWT airfoil and considering the corresponding induction factors, the main angles associated are also depicted. The aerodynamic forces (lift and drag) are decomposed into radial and tangential directions. For the present study, the values of the induction factors corresponding to each of the two airfoil sections studied, sections 54 and 81, were obtained from \cite{zahle2014comprehensive}.

\begin{figure}
    \centering
    \begin{subfigure}[t]{0.7\textwidth}
        \centering
        \includegraphics[width=\textwidth]{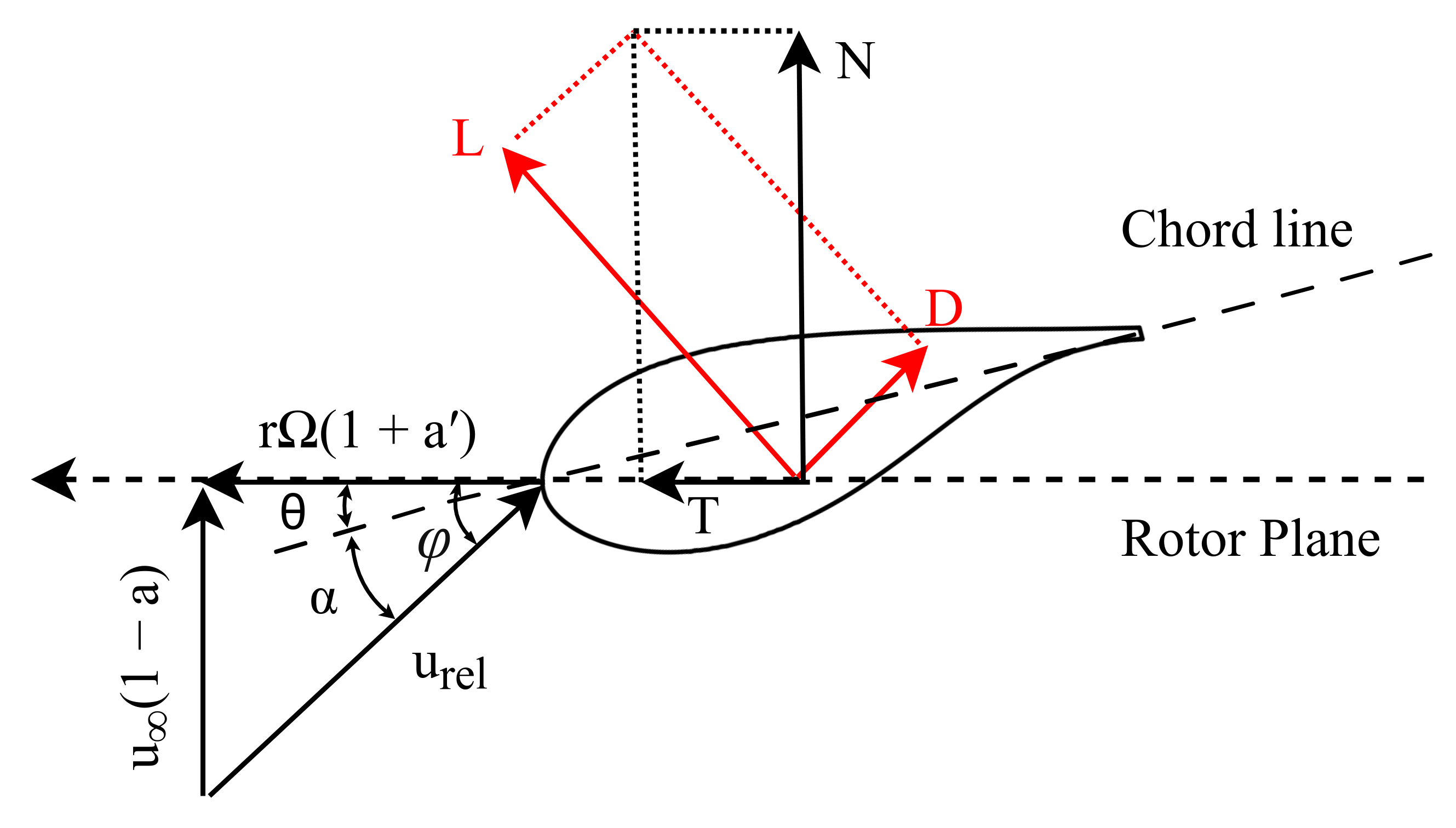}
    \end{subfigure}
    \caption{Wind and rotor velocities and forces on a generic blade section.}
    \label{fig:wind_rotor_velo}
\end{figure}

Additionally, for the computation of essential metrics such as net force, torque, angle of attack, and relative wind speed for each blade section, as described by Formulas \ref{eq:Urel}, \ref{eq:AoA}, \ref{eq:Q2}, \ref{eq:T2} from  \cite{BEMllibre,BEMtreball}, the framework necessitates the establishment of two distinct angles. These angles serve as pivotal factors in the analytical processes.
The initial angle, referred to as $\theta$, corresponds to the angle formed between the section chord and the plane of rotation, designated as the section twist angle. This angle, which is different for each section is defined in the design process of each HAWT and cannot be modified. In fact, and in order to obtain the maximum power for each given wind speed, HAWT traditionally uses the blade pitch angle, not depicted in Figure \ref{fig:wind_rotor_velo}, note that the pitch angle is constant for all blade sections and just depends on the wind speed. For the present DTU10MW RWT and when the wind speed is of $10m/s$ the pitch angle is zero.
Lastly, the other angle which plays a crucial role in the calculations is labeled as $\varphi$ and characterizes the inflow angle. This angle quantifies the divergence between the relative wind speed of each blade section and the plane of rotation, as delineated in Figure \ref{fig:wind_rotor_velo}. Notice that the Angle of Attack (AoA) is given as $\alpha=\varphi-\theta$. 

Based on the information just presented, the relation between the relative velocity $u_{rel}$, the free stream velocity $u_{\infty}$ and the tangential one $\Omega r$, $r$ being the section (airfoil) radius and $\Omega$ stands for the wind turbine turning speed, is given as \ref{eq:Urel}.

\begin{equation}
    u_{rel} = \sqrt{\left ( \Omega r \left ( 1+a' \right ) \right )^2+\left ( u_\infty\left ( 1-a \right ) \right )^2}
    \label{eq:Urel}
\end{equation}

The equation defining the angle of attack (given in degrees) for any blade section takes the form:
\begin{equation}
    \alpha = \arctan{\left ( \frac{u_{rel}\left ( 1-a \right )}{\Omega r\left ( 1+a' \right )} \right )}\cdot \frac{360}{2\pi}-\theta
    \label{eq:AoA}
\end{equation}
For a given section, the resultant force acting on the rotor plane direction $(T)$ and the force acting in a plane perpendicular to it $(N)$, are expressed as.
\begin{equation}
    T=\frac{1}{2} C \rho u_{rel}^2\left (C_l\sin \varphi -C_d\cos \varphi \right ) 
    \label{eq:Q2}
\end{equation}

\begin{equation}
    N=\frac{1}{2} C \rho u_{rel}^2\left (C_l\cos \varphi +C_d\sin \varphi\right ) 
    \label{eq:T2}
\end{equation}
From equation \ref{eq:Q2}, several insights can be derived regarding the significance of lift and efficiency across different parts of the blade: when dealing with high inflow angles at the root region (close to $90\deg.$), the contribution of the lift coefficient becomes more significant than that of drag in torque generation. Conversely, at the tip, the drag coefficient (and hence the aerodynamic efficiency) plays a more prominent role, making it crucial to enhance efficiency in this region.
Based on the Angles of Attack (AoA) associated to each section and the aerodynamic data provided by reference \cite{web:DTUReport}, it is estimated that approximately $40\%$ of airfoils will experience separated flow at a wind speed of $10m/s$, making it a good regime for enhancing the aerodynamic capabilities via AFC implementation. It should be noted that other wind velocities also have a high number of airfoils in post-stall conditions, which opens a door into improving the WT efficiency and power generated.

\section{Results} \label{results}

\subsection{Baseline Case Results Summary} \label{baseline}

Delving into the results of the baseline case simulations, it will initially be presented the numerical values encompassing the time averaged lift $C_{l}$ and drag $C_{d}$ coefficients, the peak-to-peak (PTP) variations of these parameters $PTP_{C_{l}}$,  $PTP_{C_{d}}$, the aerodynamic efficiency $\eta$, the non dimensional boundary layer separation points $x_{\text{sep}}/C$ and the dimensional vortex shedding frequencies $f_0$. For the specific numerical values already mentioned, the reader should refer to Table \ref{tab:SummaryBCSresults}.

From the data presented in this table, several key insights can be gleaned about the behavior of the blade's middle sections at moderate wind speeds. It is important to highlight that the scope of simulation encompasses just two sections. Consequently, drawing comprehensive conclusions about the complete flow behavior within this blade region at this specific wind speed becomes challenging. However, despite this limitation, certain observations can be made. Notably, both of these sections are situated within the middle region of the blade. In this context, it becomes evident that they share common traits such as a heightened lift-to-drag ratio and rather small PTP oscillation magnitudes. Furthermore, it is noteworthy that the oscillation frequencies remain relatively subdued for Section 54, whereas for Section 81 and due to the fact that there is no boundary layer separation, small vortical structures are just appearing at the airfoil trailing edge, the oscillations occur at a much more rapid rate.

\begin{table}[]
    \centering
    \begin{tabular}{ccccccccc}
    \hline
    \textbf{Wind speed} & \textbf{Section num.} & $\mathbf{C_{l}}$ & $\mathbf{C_{d}}$ & ${\eta=(C_l/C_d)}$ & $\mathbf{PTP_{C_{l}}}$ & $\mathbf{PTP_{C_{d}}}$ & 
    $\mathbf{x_{\text{sep}}/c}$ & $\mathbf{f_0 (Hz)}$ \\ \hline
    \multirow{2}{*}{$\mathbf{10\:m/s}$} & 54 & 1.5623 & 0.10625 & 14.71 & 0.281 & 0.104 & 0.423 & 7.1 \\
    & 81 & 1.50973 & 0.04464 & 33.82 & 0.0038 & 0.0011 & - & 28.8 \\ \hline
    \end{tabular}
\caption{Summary of the main parameters of the baseline cases simulations. Including the averaged aerodynamic coefficients, the PTP values, the separation point and the vortex shedding frequency.}
\label{tab:SummaryBCSresults}
\end{table}

\begin{figure}
    \centering
     \begin{subfigure}[b]{1\textwidth}
         \centering
         \includegraphics[width=0.9\textwidth]{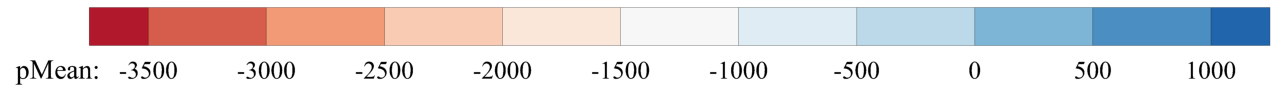}
         \label{subfig:CpS54}
     \end{subfigure}
     \hfill
     \begin{subfigure}[b]{1\textwidth}
         \centering
         \includegraphics[width=0.9\textwidth]{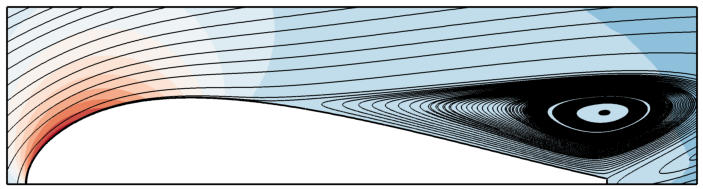}
         \caption{Section 54}
         \label{subfig:CfS54}
     \end{subfigure}
     \par\bigskip 
     \begin{subfigure}[b]{1\textwidth}
         \centering
         \includegraphics[width=0.9\textwidth]{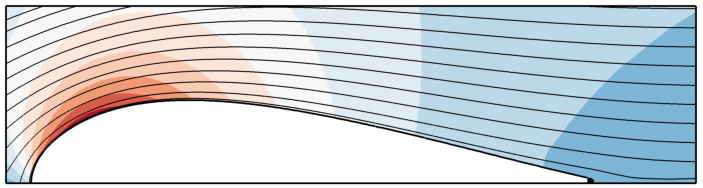}
         \caption{Section81}
         \label{subfig:FFTS54}
     \end{subfigure}
     \hfill
    \caption{Velocity fields given as time averaged streamlines and time averaged pressure fields for sections 54 and 81 }
    \label{fig:7results}
\end{figure}

In order to have an illustrative overview of the data presented in Table \ref{tab:SummaryBCSresults}, Figure \ref{fig:7results} provides the time averaged of the streamlines at the upper surface of the two airfoils studied as well as the time averaged of the pressure fields. The location where the fluid has a maximum acceleration, the area where time averaged pressure is minimum, is almost the same for both sections. The vortical structures forming on top of section 54 and its associated boundary layer separation point can now be clearly seen. The very small vortical structures having a frequency of around $28.8 Hz$ are not spotted form the view presented of section 81. Yet this information is obtained from the Fast Fourier Transformation (FFT) of the dynamic lift signal. 

Delving into the post-processing graphs, Figures \ref{subfig:CpS54},\ref{subfig:CfS54} present the pressure and friction coefficients comparison for the two sections studied, from where it is seen that a high lift coefficient is directly linked to a large pressure coefficient below the airfoil and a particularly low pressure coefficient on the airfoil upper surface. In section 54, there is a pronounced suction peak on the upper surface near the leading edge, followed by a rapid pressure recovery and subsequently, a nearly constant value of $C_p$. This observation suggests a potential flow separation near that particular location. The friction coefficient becomes negative once the boundary layer is separated, and the pressure on the airfoil upper surface approaches the atmospheric one. The pressure coefficient distribution for Section 81 also includes a high suction peak at early chord coordinates followed by a smooth pressure recovery which appears to point out that no flow separation is to be expected. The friction coefficient also shows a smooth decrease along the airfoil chord, but does not becomes negative.

\begin{figure}
     \centering
     \newcommand{\subfigurehspace}{.02\linewidth}
     \setlength{\belowcaptionskip}{0.2\baselineskip}
     \begin{subfigure}[b]{0.48\textwidth}
         \centering
         \includegraphics[width=1\textwidth]{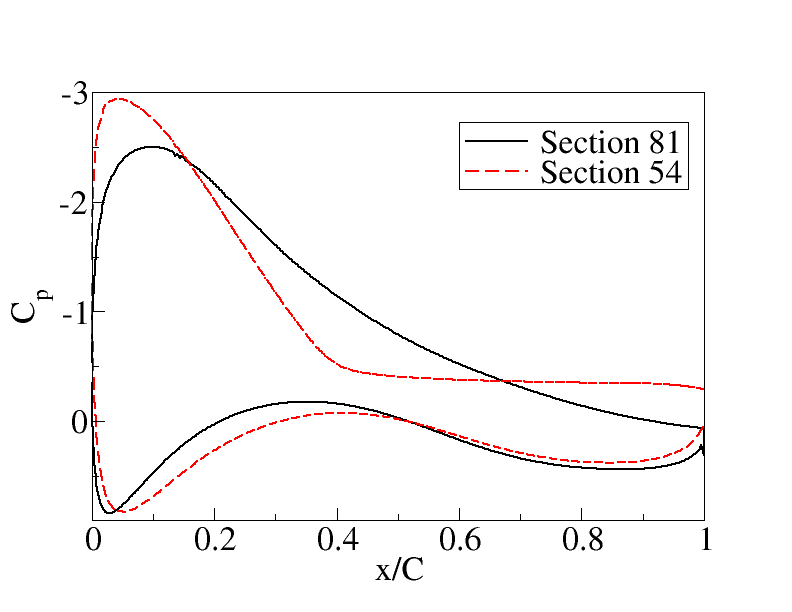}
         \caption{}
         \label{subfig:CpS54}
     \end{subfigure}
     \hfill
     \begin{subfigure}[b]{0.48\textwidth}
         \centering
         \includegraphics[width=1\textwidth]{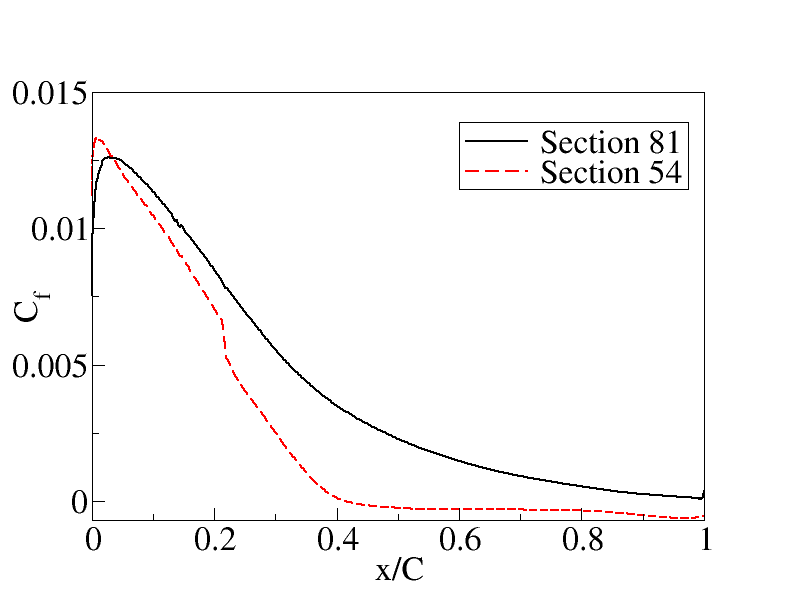}
         \caption{}
         \label{subfig:CfS54}
     \end{subfigure}
     \par\bigskip 
     \begin{subfigure}[b]{0.49\textwidth}
          \centering
          \includegraphics[width=1\textwidth]{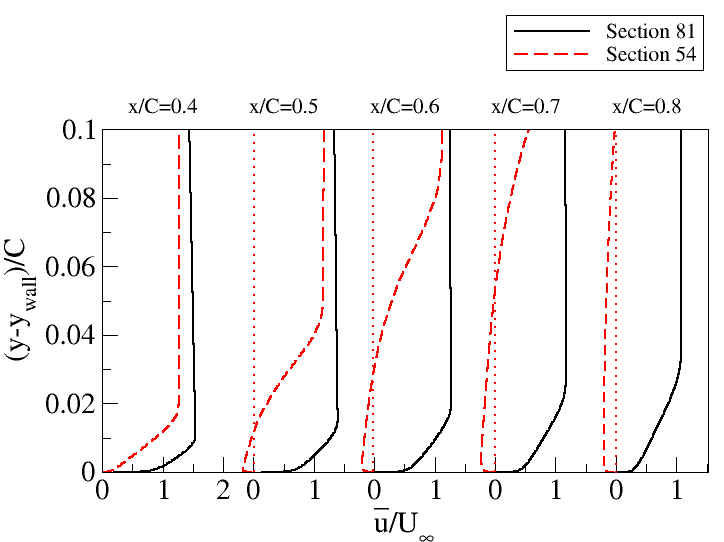}
          \caption{}
          \label{subfig:BLUS54}
      \end{subfigure}
     \begin{subfigure}[b]{0.49\textwidth}
         \centering
         \includegraphics[width=1\textwidth]{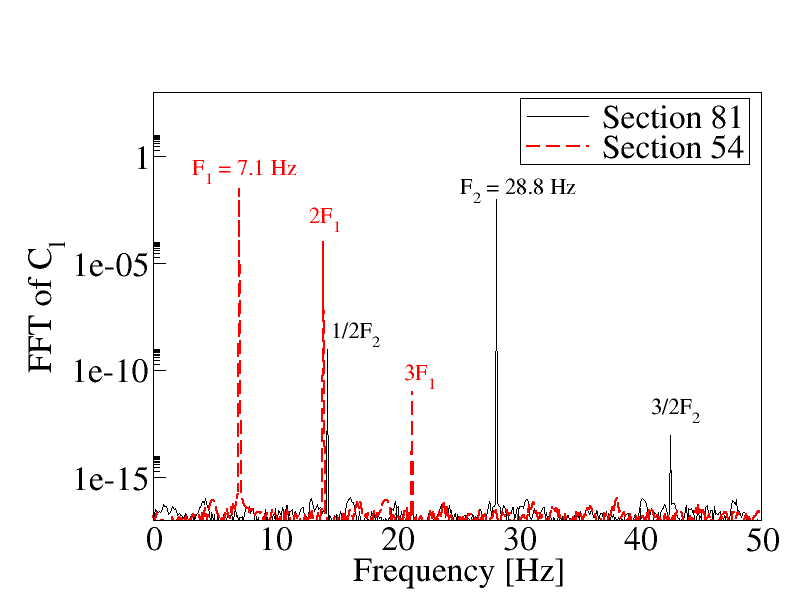}
         \caption{}
         \label{subfig:FFTS54}
     \end{subfigure}
     \hfill
    \caption{Post-processing results for sections 54 and 81. Pressure coefficient distribution along upper and lower airfoils surfaces, (a). Tangential friction coefficient distribution along airfoils upper surfaces, (b). Mean velocity distribution tangent to the airfoils upper surfaces and as a function of the normalised wall normal distance, (c). Fast Fourier Transform (FFT) of lift coefficients, (d).}
    \label{fig:S54results}
\end{figure}

The velocity field tangent to the airfoil surface as well as the FFT transformation of the dynamic lift coefficient for both sections are introduced in Figures \ref{subfig:BLUS54} and \ref{subfig:FFTS54}, respectively.
As discussed previously, when the friction coefficient reaches a null value at a certain point along the airfoil surface, it indicates flow separation. This phenomenon is illustrated in Figure \ref{subfig:CfS54}, where the friction coefficient of Section 54 is observed to intersect the 0 value at an $x$ coordinate of $2.566\:m$ from the leading edge. In contrast, the $C_f$ of section 81 is not showing any $x$ axis crossing, indicating there is no flow separation. The boundary layer separation point is further seen in Figure \ref{subfig:BLUS54}, where the velocity distribution tangential to the airfoil surface is shown along the chord, notice that the velocity field is already negative nearby the wall at a position $x/c=0.5$ and further downstream, as previously observed in Figure \ref{subfig:CfS54}. Finally, Figure \ref{subfig:FFTS54} presents the vortex shedding frequency read from the lift coefficient, notice that as expected, when the boundary layer is separated the vortex shedding frequency associated is much smaller, section 54, than when the separation occurs at the airfoil trailing edge, section 81.

\subsection{Airfoil section chosen for the AFC implementation} \label{sec:AFC}

From the two sections analyzed just section 54 presents flow separation, this section is therefore chosen for AFC implementation via a parametric analysis. AFC is capable of reducing drag and increasing lift forces, its implementation shall generate a reduction in aerodynamic loads on the blade's downstream direction while increasing the torque generated.

The placement of the AFC groove is based on previous research works \cite{feero2017influence, amitay2001aerodynamic,amitay2002role}, where it was concluded that the optimum location for the groove was in the vicinity of the boundary layer separation point. 
Which for the chosen section occurs at $x=2.566\:m$, $x/C=0.423$ (approximately $42\:\%$ of the airfoil's chord).
As defined in equation \ref{eq:Cmu}, the groove width is directly related with the jet momentum coefficient. For the present study, a width of $h= 0.5\:\%$ of the airfoil chord was selected, dimension widely accepted in the literature \cite{tousi2022large,tadjfar2020optimization,couto2022aerodynamic}.
In the present parametric study, jet position and jet width were kept constant across the entire set of CFD simulations, therefore the remaining three AFC parameters, momentum coefficient, jet inclination angle and jet pulsating frequency, will be optimized in this research. 

\textcolor{black}{To perform the parametric study, it is needed to modify the existing baseline mesh of Section 54, implementing a slot in the upper surface of the airfoil upwind of the separation point ($x/c=0.4$), as depicted in Figure \ref{subfig:MeshAFCdet}. A zoomed view of the mesh implemented in the vicinity of the jet groove has been previously introduced in Figure \ref{subfig:MeshAFCgen}.}

\subsection{Momentum coefficient $C_{\mu}$ optimization.}
 \label{sec:results_1}

Synthetic Jet Actuators (SJA) generate zero net mass flux, which means that the mass flow through the AFC groove must be equal during both the suction and blowing phases. In order to achieve this statement, the time dependent velocity profile for the AFC simulations is often chosen to be a sinusoidal trigonometric function, which is usually expressed as:

\begin{equation}
     u_{\text{jet}}(t)=u_{\max}\cdot \sin\left ( 2\pi f \cdot t \right )
     \label{eq:usin}
\end{equation}

Where $f$ is the synthetic jet actuation frequency, $u_{\max}$ is the maximum SJ outgoing/incoming velocity and $t$ is the time. 

Based on previous studies, \cite{tuck2008separation,kitsios2011coherent,buchmann2013influence,zhang2015direct,kim2009separation,monir2014tangential}, for optimum performance the synthetic jet actuation frequency should be the same or a multiple/sub-multiple of the vortex shedding one. This is why in our parametric analysis we will initially set the SJ pulsating frequency constant and equal to the vortex shedding one, which was $f_0=7.1\:Hz$.
The second parameter which will be initially kept constant is the jet inclination angle. A reasonable initial guess would be to inject the flow almost tangentially to the airfoil surface, taking advantage of the Coanda effect \cite{book:Coanda}. Thus, by injecting the pulsating flow tangentially to the airfoil surface, the Coanda effect can help to maintain the incoming flow attached to the airfoil. Based on several previous studies \cite{Tousi2021,tadjfar2020optimization}, we have decided to start the parametric analysis maintaining constant at 10 degrees the SJ injection angle. At this point, four AFC parameters are initially set and the parametric optimization starts via evaluating the effect of the momentum coefficient on the flow dynamics. The relation of the maximum velocity amplitude ($u_{\max}$) and the momentum coefficient can be extracted from \ref{eq:Cmu}. According to \cite{momcoeff}, relatively low momentum coefficients for AFC at high Reynolds numbers can lead to improvements in the aerodynamic forces of airfoils. Kim and Kim \cite{kim2009separation}, numerically investigated flow separation control with SJAs on a NACA23012 airfoil at $Re=2.19\cdot 10^{6}$ and they found that as momentum coefficient increased, lift improvement in comparison to the one obtained at small $C_{\mu}$ decreased. Moreover, to prevent reaching compressible flow effects on the groove exit, $C_\mu$ should be kept low. Therefore, momentum coefficients ranging between $C_\mu=0.0001$ and $C_\mu=0.01$ have been selected as an initial guess for the present application.  

Table \ref{tab:CmuIterations} presents a comprehensive overview of the AFC simulations for all $C_{\mu}$ evaluated as well as for the baseline case, offering numerical insights into the averaged lift and drag coefficients, aerodynamic efficiency, and peak-to-peak values for lift and drag coefficients. Notably, the data reveals that the momentum coefficient yielding the highest lift force is $C_\mu = 0.007$, the drag being particularly small under these conditions. Significantly, it is observed a reduction in the amplitude of lift and drag coefficient oscillations when compared to the baseline case. It's worth noting that for the iterative process involving momentum coefficient, this reduction isn't exceptionally pronounced, however, it can be deduced that, especially for high lift coefficients and low drag coefficients (i.e. high efficiencies), there is a slight decrease in the PTP values when compared to lower efficiency scenarios. In conclusion, it can be affirmed that delaying the boundary layer results in lift enhancement and drag reduction, accompanied by a dampening of the fluctuations of lift and drag forces.

\begin{table}[]
\centering
\renewcommand{\arraystretch}{1.2}
\setlength{\tabcolsep}{15pt}
\begin{tabular}{cccccc}
\hline
$C_{\mu}$ & $C_{l}$ & $C_{d}$ & $\eta$ & $PTP_{C_{l}}$ & $PTP_{C_{d}}$ \\
\hline
0.0001 & 1.947 & 0.073 & 26.678 & 0.122 & 0.047 \\
0.0003 & 1.926 & 0.087 & 22.169 & 0.154 & 0.063 \\
0.0005 & 2.000 & 0.089 & 22.550 & 0.244 & 0.095 \\
0.0007 & 2.006 & 0.086 & 23.279 & 0.237 &  0.091 \\
0.001 & 2.203 & 0.068 & 32.299 & 0.222 &  0.082 \\
0.003 & 2.305 & 0.050 & 46.342 & 0.188 & 0.075  \\
0.005 & 2.352 & 0.048 & 48.982 & 0.160 &  0.064 \\
{\color[HTML]{EE2D04} 0.007} & {\color[HTML]{EE2D04}2.450} & {\color[HTML]{EE2D04} 0.050} & {\color[HTML]{EE2D04} 48.712} & {\color[HTML]{EE2D04} 0.144} & {\color[HTML]{EE2D04} 0.046}  \\
0.01 & 2.269 & 0.069 & 33.084 & 0.086 &  0.039 \\ \hline
\textbf{BCS S54} & 1.562 & 0.106 & 14.71 & 0.281 & 0.104 \\ \hline
\end{tabular}
\caption{ Time averaged $C_l$, $C_d$, $\eta$ and peak to peak  $PTP_{C_{d}}$, $PTP_{C_{d}}$ aerodynamic values obtained for the different $C_{\mu}$ considered. Jet inclination angle ($\theta_{\text{jet}}=10\: deg.$), forcing frequency ($F^+=1$), groove width $h=0.005C$ and groove position $x/C=0.4$. For comparison, the baseline case values are presented in the last row.}
\label{tab:CmuIterations}
\end{table}

To gain further insight into the relationship between the lift coefficient and the momentum coefficient, Figure \ref{subfig:MomIT_cl} is presented. The graph illustrates that at very low momentum coefficients, the jet fails to inject sufficient energy into the boundary layer, and the BL is likely to remain separated. As the momentum coefficient increases, the jet becomes more effective in energetically injecting air, a delay in the boundary layer separation is to be expected under these conditions. However, at excessively high momentum coefficients where the jet velocity is very high, boundary layer tripping is likely to occur, the boundary layer separation is promoted. The drag coefficient graph, Figure \ref{subfig:MomIT_cd}, reveals a minimum value at $C_\mu=0.005$, but it is followed by a slightly larger $C_d$ value for $C_\mu=0.007$. In terms of aerodynamic efficiency, Figure \ref{subfig:MomIT_E} demonstrates that the maximum efficiency is achieved at $C_\mu=0.005$. This reinforces the selection of the momentum coefficient that yields maximum lift ($C_\mu=0.007$), as it also exhibits a high efficiency, almost the same as the one obtained with $C_\mu=0.005$. Note that the respective peak to peak values for $C_l$ and $C_d$ are as well implemented in Figure \ref{fig:MomIT}. 

\begin{figure}[]
     \centering
     \newcommand{\subfigurehspace}{.02\linewidth}
     \setlength{\belowcaptionskip}{0.2\baselineskip}
     \begin{subfigure}[b]{0.32\textwidth}
         \centering
         \includegraphics[width=\textwidth]{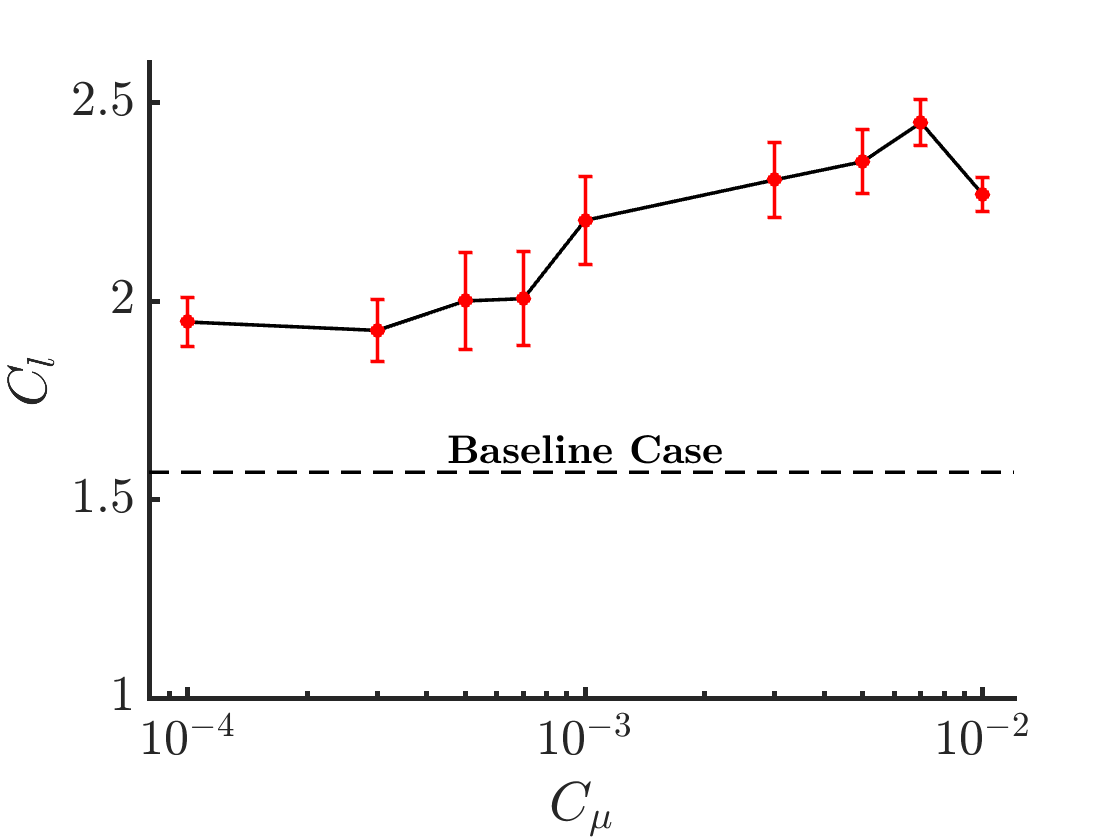}
         \caption{}
         \label{subfig:MomIT_cl}
     \end{subfigure}
     \hfill
     \begin{subfigure}[b]{0.32\textwidth}
         \centering
         \includegraphics[width=\textwidth]{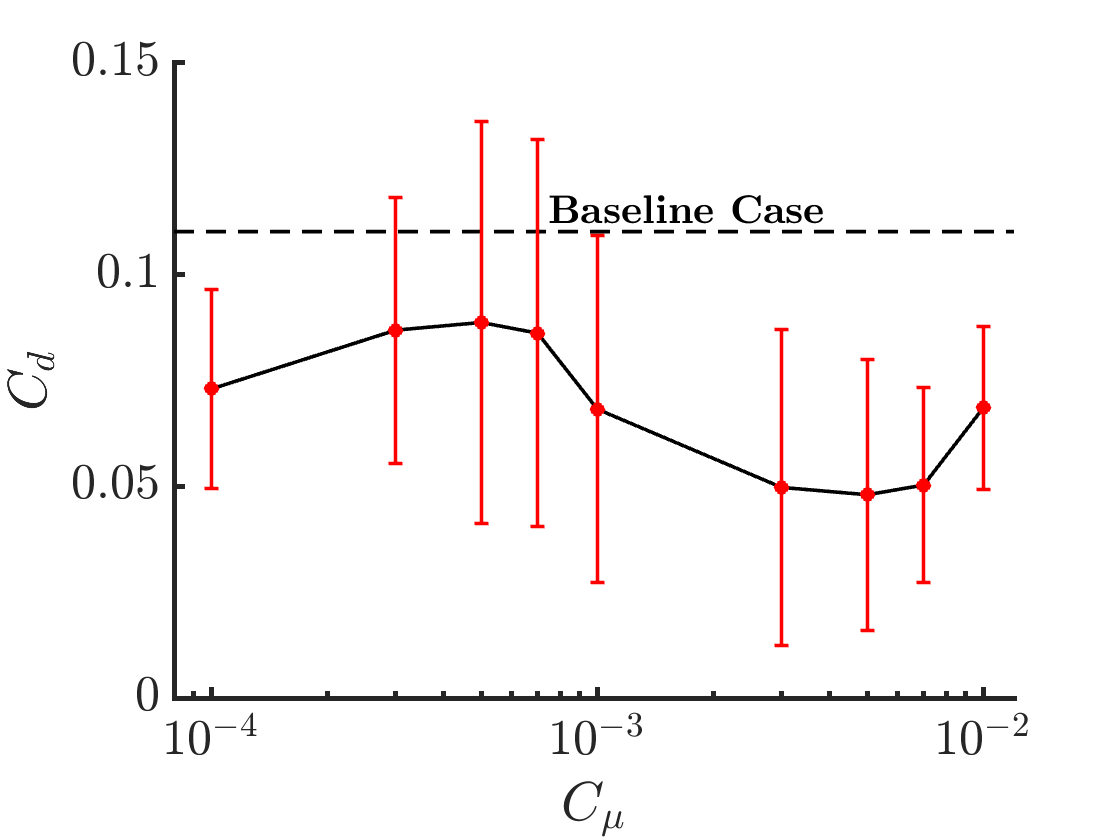}
         \caption{}
         \label{subfig:MomIT_cd}
     \end{subfigure}
     \hfill 
     \begin{subfigure}[b]{0.32\textwidth}
         \centering
         \includegraphics[width=\textwidth]{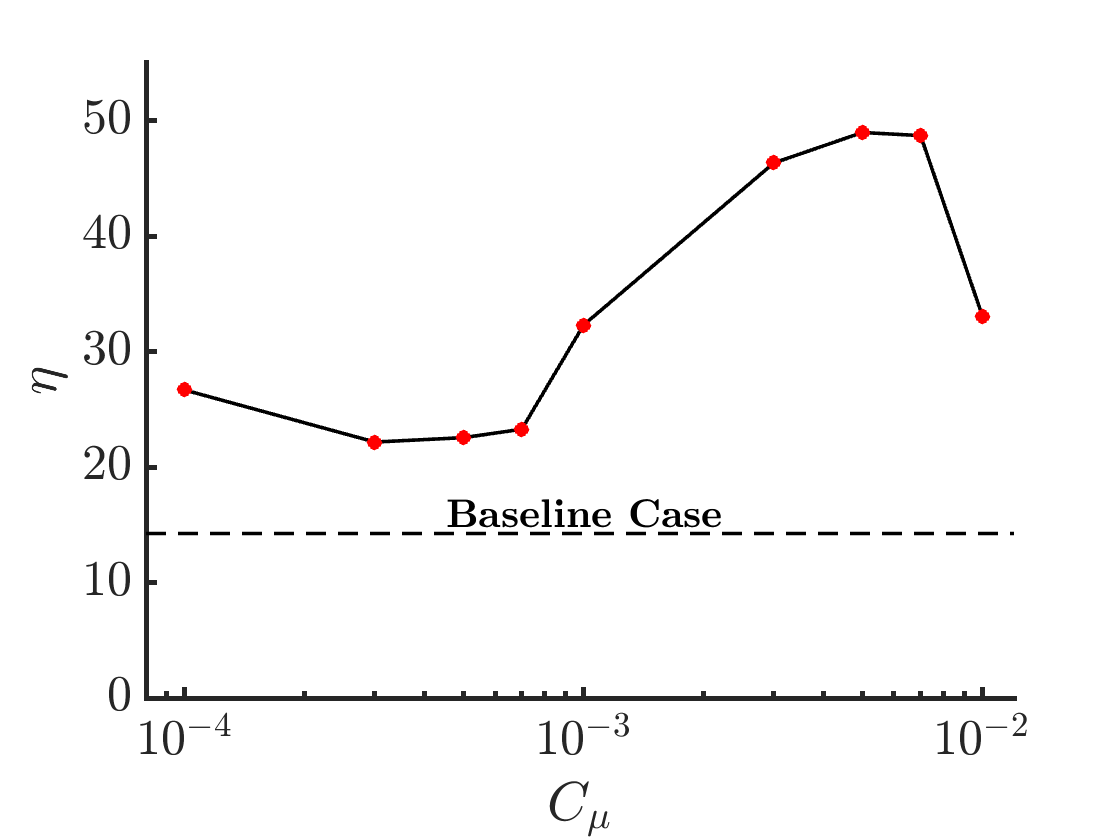}
         \caption{}
         \label{subfig:MomIT_E}
     \end{subfigure}
        \caption{Lift coefficient (a), drag coefficient (b) and aerodynamic efficiency (c) variation as a function of momentum coefficient, the vertical bars characterize the corresponding peak to peak amplitude. Constant jet angle ($\theta_{\text{jet}}=10\: deg.$) and forcing frequency ($F^+=1$).}
        \label{fig:MomIT}
\end{figure}

 \subsection{Jet injection angle optimization}
 \label{sec:results_1}

 To optimize the jet inclination angle ($\theta_{\text{jet}}$), the momentum coefficient obtained in the previous subsection ($C_\mu=0.007$) will be kept constant, along with SJ pulsating frequency ($f_0=7.1\:Hz$) ($F^+=1$), the groove position ($x/C=0.4$) and width ($h=0.5\% C$). Ten different jet inclination angles ranging between $2$ and $50$ degrees were simulated, Table \ref{tab:angleIterations} presents the time averaged values for the aerodynamic coefficients and their respective peak to peak amplitudes corresponding to each jet angular position evaluated. From this data, it is highlighted that the optimal angle obtained from the parametric optimization is $\theta_{\text{jet}}=5\:deg.$, being its associated efficiency of $\eta=41.067$, however, it should be mentioned that $\theta_{\text{jet}}=2\:deg.$ yields a higher lift coefficient although the efficiency drastically decays to $\eta=16.151$. Concerning the oscillations of the lift and drag coefficients in relation to the jet angle, it can be observed that the lift coefficient's PTP values remain relatively stable for injection angles below 25 degrees. Furthermore, in the case of the drag coefficient, oscillations decrease notably within the region of minimum drag, specifically within the range of jet angles where $5\leq\theta_{\text{jet}}\leq 20$. The observed reduction in PTP values for both lift and drag coefficients signifies an improvement in the dynamic forces, which reduce vibration when compared to the baseline case.

 \begin{table}[]
\centering
\renewcommand{\arraystretch}{1.2}
\setlength{\tabcolsep}{15pt}
\begin{tabular}{cccccc}
\hline
$\theta_{\text{jet}}$ & ${C_{l}}$ & ${C_{d}}$ & $\eta$ & ${PTP_{C_{l}}}$ & ${PTP_{C_{d}}}$ \\
\hline
2 & 2.633 & 0.163 & 16.151 & 0.163 & 0.087  \\
{\color[HTML]{EE2D04} 5} & {\color[HTML]{EE2D04} 2.537} & {\color[HTML]{EE2D04} 0.062} & {\color[HTML]{EE2D04} 41.067} & {\color[HTML]{EE2D04} 0.090} & {\color[HTML]{EE2D04} 0.050} \\
7 & 2.490 & 0.057 & 43.510 & 0.085 & 0.041  \\
10 & 2.450 & 0.050 & 48.712 & 0.114  & 0.046  \\
15 & 2.345 & 0.052 & 45.297 & 0.092 & 0.035  \\
20 & 2.249 & 0.057 & 39.114 & 0.110 & 0.036  \\
25 & 2.225 & 0.066 & 33.505 & 0.110 & 0.070  \\
30 & 2.242 & 0.063 & 35.466 & 0.075 &  0.028 \\
40 & 2.033 & 0.082 & 24.795 & 0.309 & 0.111  \\
50 & 2.079 & 0.070 & 29.592 & 0.095 & 0.036  \\ \hline
\textbf{BCS S54} & 1.562 & 0.106 & 14.71 & 0.281 & 0.104 \\ \hline
\end{tabular}
\caption{Time averaged aerodynamic coefficients $C_l$, $C_d$, $\eta$, and their respective peak to peak values $PTP_{C_{l}}$ and $PTP_{C_{d}}$, for a set of different jet inclination angles $\theta_{jet}$. Momentum coefficient ($C_\mu=0.007$), forcing frequency ($F^+=1$), groove position ($x/C=0.4$) and width ($h=0.5\% C$), are kept constant.}
\label{tab:angleIterations}
 \end{table}

\begin{figure}
     \centering
     \newcommand{\subfigurehspace}{.02\linewidth}
     \setlength{\belowcaptionskip}{0.2\baselineskip}
     \begin{subfigure}[b]{0.32\textwidth}
         \centering
         \includegraphics[width=\textwidth]{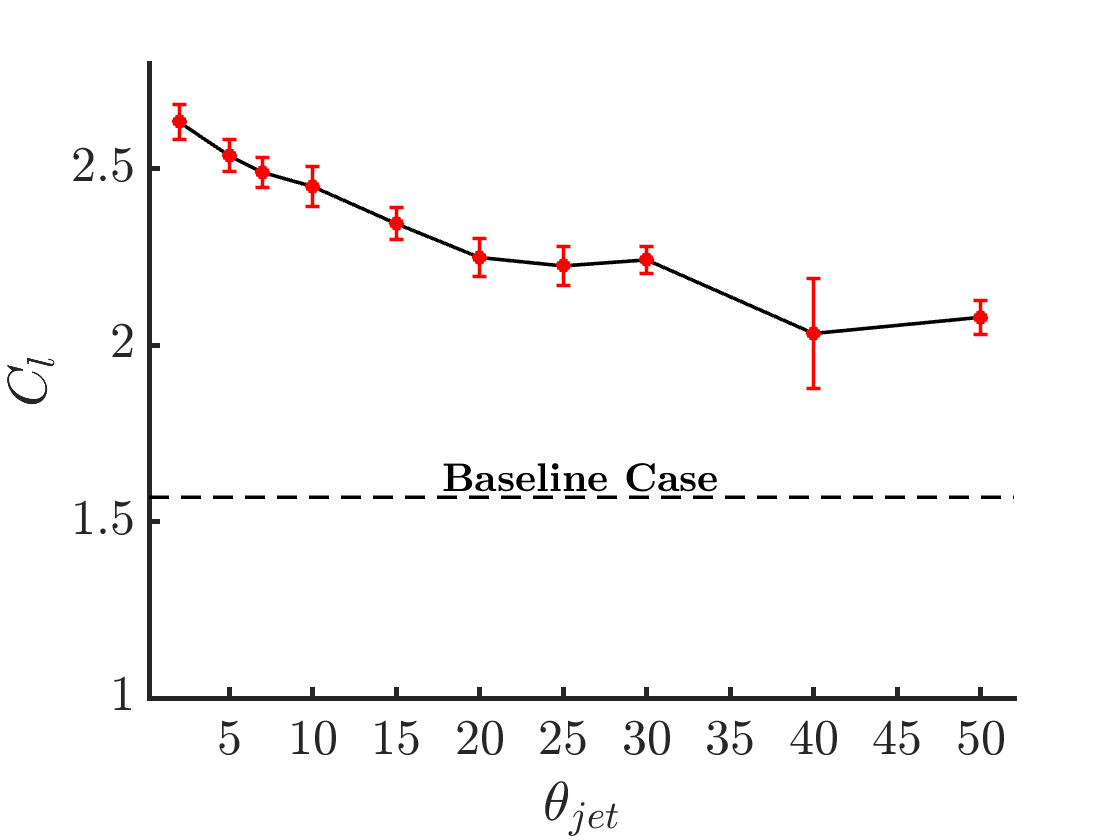}
         \caption{}
         \label{subfig:AngleIT_cl}
     \end{subfigure}
     \hfill
     \begin{subfigure}[b]{0.32\textwidth}
         \centering
         \includegraphics[width=\textwidth]{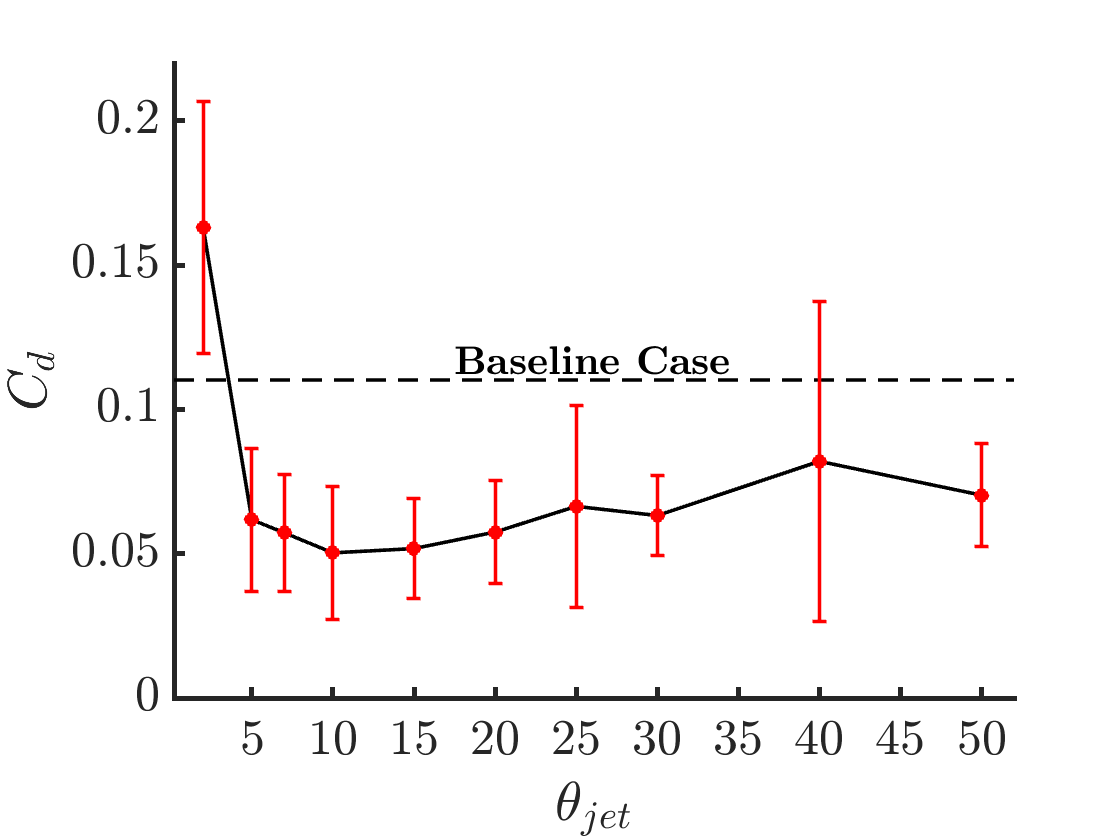}
         \caption{}
         \label{subfig:AngleIT_cd}
     \end{subfigure}
     \hfill 
     \begin{subfigure}[b]{0.32\textwidth}
         \centering
         \includegraphics[width=\textwidth]{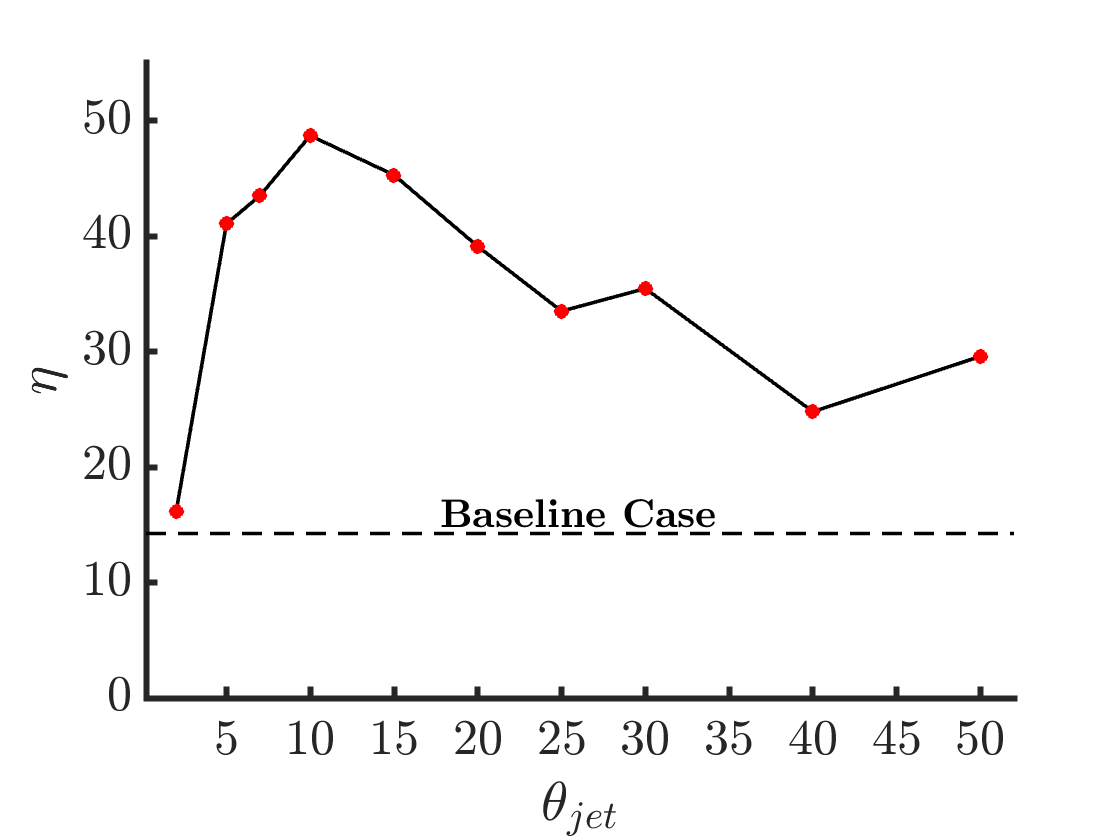}
         \caption{}
         \label{subfig:AngleIT_E}
     \end{subfigure}
        \caption{Lift coefficient (a), drag coefficient (b) and aerodynamic efficiency (c) variation as a function of the jet angle, the vertical bars characterize the corresponding peak to peak amplitude. Constant momentum coefficient ($C_\mu=0.007$) and forcing frequency ($F^+=1$).}
        \label{fig:AngleIT}
\end{figure}

When exploring the lift coefficient as a function of the jet angle in Figure \ref{subfig:AngleIT_cl}, it is observed that the lift keeps increasing as the jet angle decreases, suggesting a delay of the boundary layer separation. Additionally, studies such as \cite{angle1}, \cite{angle2}, and \cite{angle3} have extensively explored the effects of tangential flow injection and have demonstrated its ability to promote flow reattachment and lift increase. The Coanda effect refers to the tendency of a fluid jet to adhere to a nearby surface due to curvature of the solid body. By decreasing the jet angle, the airflow is directed closer to the airfoil surface, effectively leveraging the Coanda effect to enhance boundary layer reattachment. Thus, and regarding the results obtained in this research, the observed increase in lift coefficient with decreasing jet angle aligns with the predictable outcomes of previous research.

When considering the choice of $\theta_{\text{jet}}=5\:deg.$ as the optimal jet angle for the AFC implementation, it is essential to examine the effects on drag coefficient and aerodynamic efficiency, Figures \ref{subfig:AngleIT_cd}, and \ref{subfig:AngleIT_E}, respectively. The drag coefficient remains relatively low for most of the studied jet angles; however, as the injection direction approaches tangential flow, the drag coefficient experiences a notable increase, even higher than the baseline case, leading to a rapid decrease in aerodynamic efficiency. Considering the airfoil efficiency Figure \ref{subfig:AngleIT_E} and Table \ref{tab:angleIterations}, it is obvious that the maximum efficiency is obtained at $\theta_{jet}=10^{\circ}$, but as explained before the final jet inclination angle chosen is based on a trade-off between maximum lift and maximum efficiency.

\subsection{Forcing frequency optimization}

At this point, the two previously optimized AFC parameters, momentum coefficient $C_\mu=0.007$ and jet angle $\theta_{\text{jet}}=5\:deg.$ are going to be held constant. Being the remaining parameter to be optimized the jet pulsating frequency $f$. Eight different non dimensional frequencies are analyzed in this part of the study, their values range between $0.1 \leq F^+ \leq 10$. The results obtained after the corresponding CFD simulations are presented in Table \ref{tab:FIterations}, where the baseline case data is introduced in the last line, for comparison. Based on the information presented, it can be concluded that the optimal value for the jet non dimensional forcing frequency is $F^+=3$. This choice is supported by the highest lift to drag ratio, excellent lift capabilities $C_{l}$ and a particularly low drag coefficient $C_{d}$.

Figure \ref{subfig:FreqIT_cl} is presenting the lift coefficient and its peak to peak amplitude for the different pulsating frequencies evaluated. The first thing to observe is that variations of the pulsating frequency are slightly modifying the lift, a small  decrease in lift is observed when considering the smallest forcing frequencies. However, once the forcing frequency reaches approximately $F^+=2$, the lift coefficient remains relatively constant suggesting that the optimal value for the forcing frequency should be greater than $2$ ($F^+\geq2$). Moreover, it is important to mention that at higher forcing frequencies, the peak-to-peak values for the lift coefficient increase significantly which indicates the presence of larger amplitude fluctuations and thus, higher structural loads.
When checking the drag coefficient values obtained at the different pulsating frequencies evaluated, Figure \ref{subfig:FreqIT_cd}, minimum drag values are observed at $F^+=0.1$ and around $F^+=3$. It therefore seems that an analysis of the aerodynamic efficiency needs to be made in order to determine the optimal forcing frequency. From Figure\ref{subfig:FreqIT_E} it can be seen a high aerodynamic efficiency at $F^+=0.1$, although it will not be chosen as the optimal value due to its lower lift coefficient. In contrast, regarding the remaining values for frequency, it is seen an efficiency peak at $F^+=3$, which offers a significant increase in lift coefficient, leading to an increase in torque generation, while maintaining a low drag coefficient (see Figure \ref{subfig:FreqIT_cd}). Additionally, the amplitude of oscillation at this frequency is relatively low, suggesting that it will have minimal impact on structural loads.

\begin{table}[]
\centering
\renewcommand{\arraystretch}{1.2}
\setlength{\tabcolsep}{15pt}
\begin{tabular}{cccccc}
\hline
${F^+}$ & ${C_{l}}$ & ${C_{d}}$ & ${\eta}$ & ${PTP_{C_{l}}}$ & ${PTP_{C_{d}}}$ \\ \hline
0.1 & 2.499 & 0.043 & 56.643 & 0.214 & 0.065  \\
0.5 & 2.524 & 0.067 & 37.856 & 0.059 & 0.057  \\
1 & 2.537 & 0.062 & 41.067 & 0.090 & 0.050  \\
2 & 2.569 & 0.050 & 51.325 & 0.218 & 0.085  \\
{\color[HTML]{EE2D04} 3} & {\color[HTML]{EE2D04} 2.571} & {\color[HTML]{EE2D04} 0.048} & {\color[HTML]{EE2D04} 53.563} & {\color[HTML]{EE2D04} 0.147} & {\color[HTML]{EE2D04} 0.063} \\
4 & 2.576 & 0.052 & 49.897 & 0.381 & 0.086 \\
5 & 2.557 & 0.056 & 45.532 & 0.269 & 0.099  \\
10 & 2.560 & 0.099 & 25.932 & 0.557 & 0.110  \\ \hline
\textbf{BCS S54} & 1.562 & 0.106 & 14.71 & 0.281 & 0.104 \\ \hline
\end{tabular}
\caption{Time averaged aerodynamic coefficients $C_l$, $C_d$, $\eta$, and their respective peak to peak $PTP_{C_{d}}$, $PTP_{C_{d}}$ amplitude, as a function of the SJ non dimensional pulsating frequency. Momentum coefficient ($C_\mu=0.007$), jet angle ($\theta_{\text{jet}}=5\: deg.$), groove position ($x/C=0.4$) and width ($h=0.5\% C$), are kept constant.} 
\label{tab:FIterations}
\end{table}

\begin{figure}
     \centering
     \newcommand{\subfigurehspace}{.02\linewidth}
     \setlength{\belowcaptionskip}{0.2\baselineskip}
     \begin{subfigure}[b]{0.32\textwidth}
         \centering
         \includegraphics[width=\textwidth]{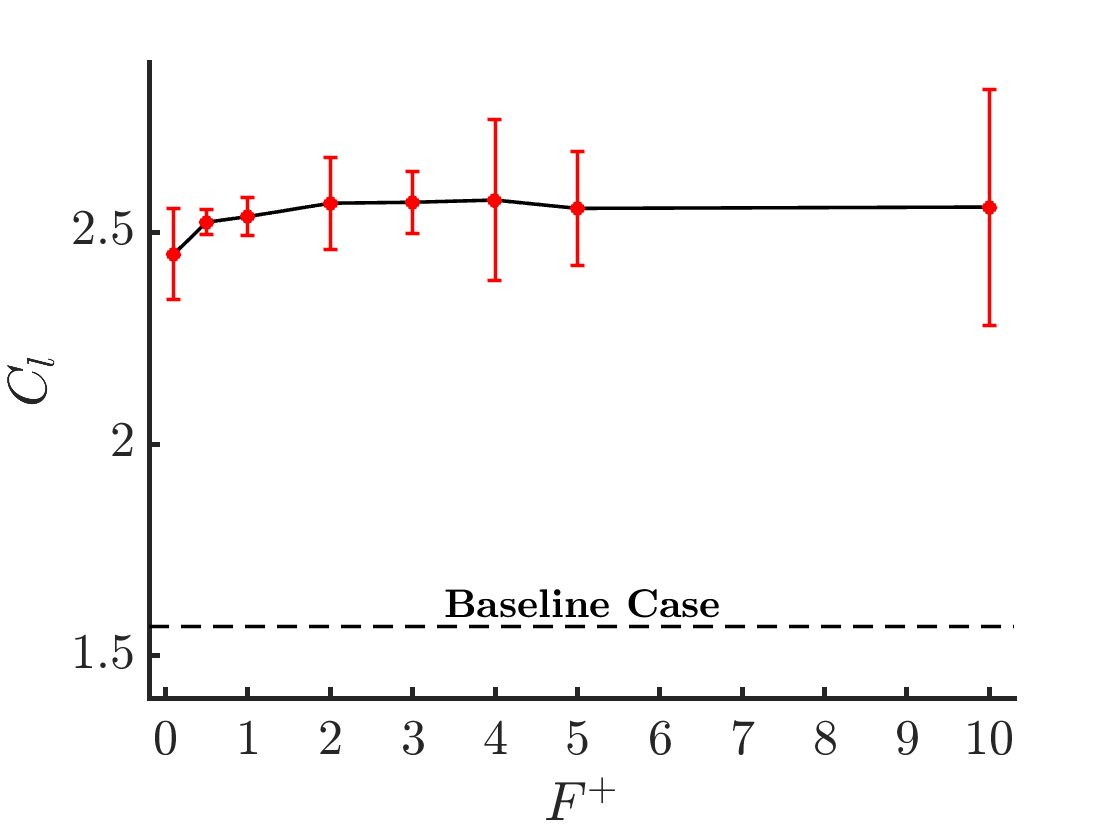}
         \caption{}
         \label{subfig:FreqIT_cl}
     \end{subfigure}
     \hfill
     \begin{subfigure}[b]{0.32\textwidth}
         \centering
         \includegraphics[width=\textwidth]{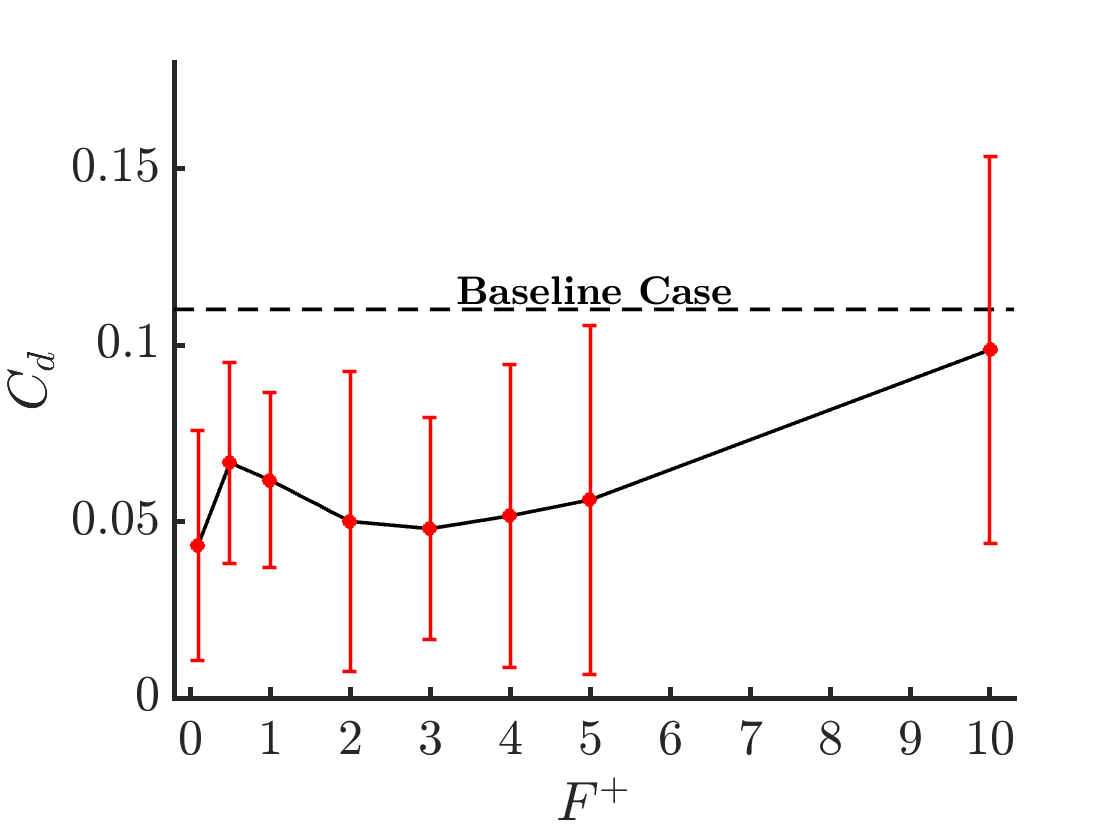}
         \caption{}
         \label{subfig:FreqIT_cd}
     \end{subfigure}
     \hfill 
     \begin{subfigure}[b]{0.32\textwidth}
         \centering
         \includegraphics[width=\textwidth]{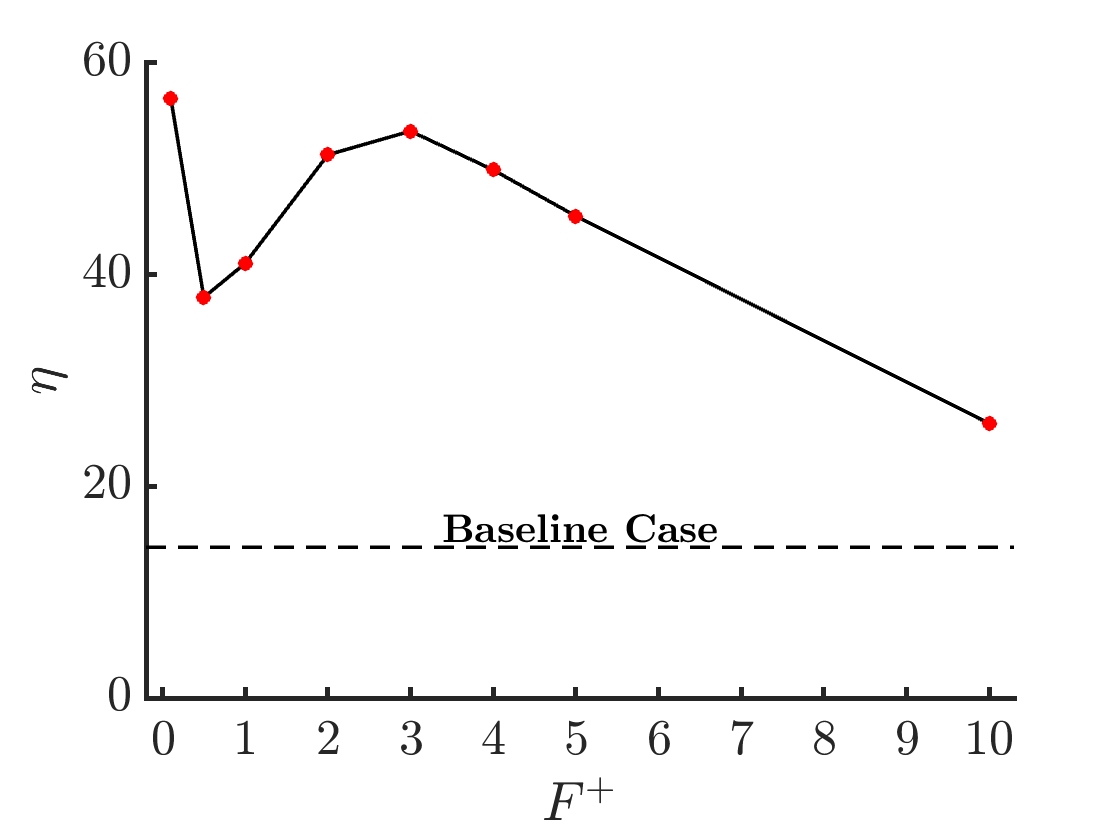}
         \caption{}
         \label{subfig:FreqIT_E}
     \end{subfigure}
        \caption{Lift coefficient (a), drag coefficient (b) and aerodynamic efficiency (c) variation as a function of forcing frequency, the vertical bars characterize the corresponding peak to peak amplitude. Constant momentum coefficient ($C_\mu=0.007$) and jet angle ($\theta_{\text{jet}}=5\: deg.$).}
        \label{fig:FreqIT}
\end{figure}

\subsection{Velocity and pressure fields for the optimal \texorpdfstring{$\theta$}{Lg}, \texorpdfstring{$C_{\mu}$}{Lg} and \texorpdfstring{$F^+$}{Lg} values} 
\label{subsec:optimum}

In the present section, the streamlines contours and the pressure fields for the baseline case and the optimum AFC parameters implementation obtained from the parametric optimization are compared in Figure \ref{fig12}. It is evident that the AFC implementation has effectively reattached the flow on the entire airfoil and so has been able to maximize its aerodynamic performance. The location of the SJA is defined in Figure \ref{fig12b}. A comparison of the pressure contour fields between the baseline and the optimum actuated case, shows a much larger low pressure area on the airfoil upper surface than the one observed for the baseline case, therefore explaining the effect of the actuation on the pressure field and the associated lift increase, which is of $64.59\%$. It is as well remarkable to highlight the lift peak to peak amplitude reduction when applying AFC, such reduction being of $47.6\%$. This considerable boost in lift performance seen in Table \ref{tab:FIterations} will be latter used for the energy assessment and torque improvement of the DTU 10MW RWT. Remarkable improvements can as well be observed in the drag coefficient when comparing the actuated case with the baseline case. Specifically, there has been a notable reduction in this parameter, amounting to an averaged decrease of $54.71\%$. Moreover, the drag peak to peak amplitude has decreased a $39.42\%$ when compared with that of the baseline-case. Such lift and drag peak-to-peak values decrease is extremely beneficial when consideing the live expectancy of any WT, then  sharply reduce the  potential vibrations that could impact the blade. To determine the effects of these oscillatory drag and lift coefficients, a comprehensive structural analysis is required, however, it is out of the scope of the present work. The aerodynamic efficiency, as shown in Table \ref{tab:FIterations}, clearly demonstrates the positive impact of the AFC implementation on the airfoil's lift-to-drag ratio. The reattachment of the flow, coupled with the increase in lift coefficient and the decrease in drag coefficient, has contributed to the mentioned significant efficiency improvement ($264.12\%$).

The non-dimensional turbulence viscosity fields, for the baseline case and the optimum AFC parameters implementation are introduced in Figure \ref{fig13}. The locations where the Reynolds stresses are particularly high, which is the airfoil trailing edge (for the AFC case) and the area where the boundary layer is separated (for the baseline case), are now clearly observed. For the case of optimum AFC parameters implementation it is now seen that small vortical structures are expected at the airfoil trailing edge, turbulence viscosity is particularly large at this point highlighting the particularly large momentum interchange between the fluid particles. It appears as if the boundary layer was about to separate just before the airfoil trailing edge.   
\begin{figure}
     \centering
     \begin{subfigure}[b]{0.8\textwidth}
         \centering
         \includegraphics[width=\textwidth]{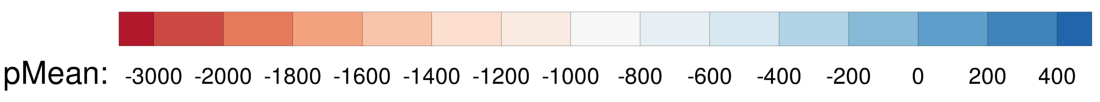}
     \end{subfigure}
     \hfill
     \begin{subfigure}[b]{0.47\textwidth}
         \centering
         \includegraphics[width=\textwidth]{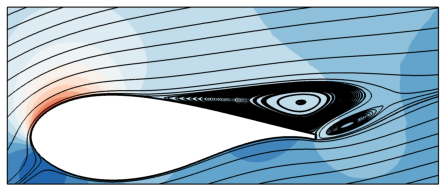}
         \caption{}
     \end{subfigure}
     \hfill 
     \begin{subfigure}[b]{0.47\textwidth}
         \centering
         \includegraphics[width=\textwidth]{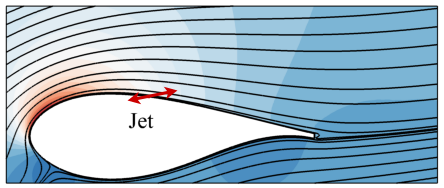}
         \caption{}
         \label{fig12b}
        \end{subfigure}
        \caption{Streamlines of temporal average velocity field and contours of averaged pressure at section 54 for the baseline case (a) and optimum AFC implementation case (b).}
    \label{fig12}
\end{figure}
\begin{figure}
     \centering
     \begin{subfigure}[b]{0.8\textwidth}
         \centering
         \includegraphics[width=\textwidth]{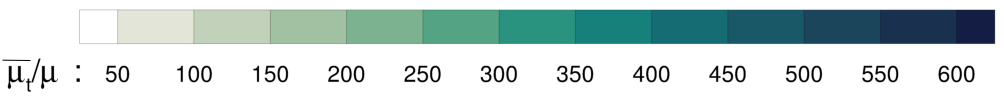}
     \end{subfigure}
     \hfill
     \begin{subfigure}[b]{0.47\textwidth}
         \centering
         \includegraphics[width=\textwidth]{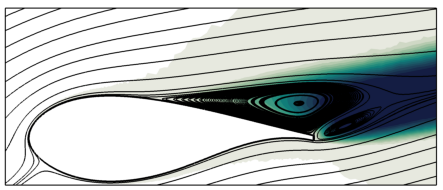}
         \caption{}
         \label{fig13a}
     \end{subfigure}
     \hfill 
     \begin{subfigure}[b]{0.47\textwidth}
         \centering
         \includegraphics[width=\textwidth]{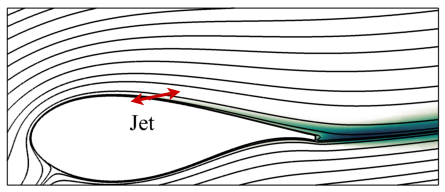}
         \caption{}
         \label{fig13b}
     \end{subfigure}
        \caption{Streamlines of temporal average velocity field and contours of non-dimensional turbulence viscosity at section 54 for the baseline case (a) and maximum lift case (b).}
  \label{fig13}
\end{figure}
\begin{figure}
     \centering
     \begin{subfigure}[b]{0.8\textwidth}
         \centering
         \includegraphics[width=\textwidth]{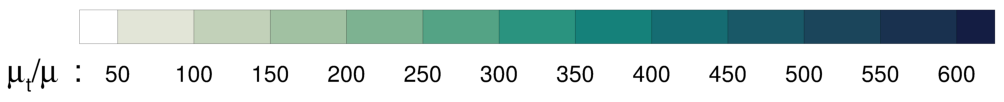}
     \end{subfigure}
     \hfill 
     \begin{subfigure}[b]{1\textwidth}
         \centering
         \includegraphics[width=\textwidth]{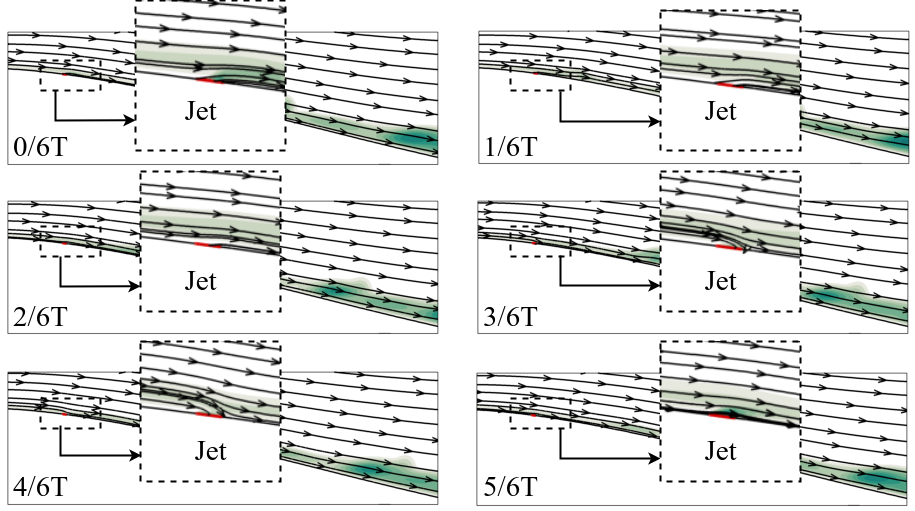}
     \end{subfigure}
        \caption{Instantaneous velocity streamlines at different phases of synthetic jet for the maximum lift case at section 54. $F^+ = 3$, $C_\mu=0.007, \theta=5^\circ$.}
  \label{fig14}
\end{figure}
In order to further clarify the effect of the synthetic jet on the flow just downstream of the groove location, the SJ oscillation period has been divided in six equal time steps and presented in Figure \ref{fig14}. At time $T=0$ the outgoing jet velocity is maximum, the turbulence viscosity is particularly high at the jet exit and downstream of it. The streamlines are suffering a particularly high curvature under these conditions. As the outgoing jet velocity decreases, (1/6)T and (2/6)T, the streamlines curvature as well as the turbulent viscosity downstream of the injection area, also decrease. When the suction phase starts (3/6)T, the streamlines just upstream of the jet location start bending inwards, being this bending curvature maximum at (4/6)T which coincides with the maximum incoming jet velocity. The turbulence viscosity around the jet location during the suction phase appears to be particularly low.    

The pressure coefficient analysis, as shown in Figure \ref{subfig:CpS54AFC}, provides insights into the behavior of the actuated and baseline cases and a comparison between the two reveals notable differences. In the actuated case, there is an evident increase in the suction peak near the leading edge in reference with the baseline case. Moreover, the pressure recovery on the upper surface of the airfoil is smooth, indicating the absence of flow separation. This is in contrast to the baseline case, which exhibits a sharp gradient of the pressure coefficient, suggesting the presence of separated flow. Notice as well that the pressure below the airfoil is higher in the AFC case than in the baseline one.   
The analysis of the tangential friction coefficient ($C_{f}$) Figure \ref{subfig:CfS54AFC}, provides valuable insights into the flow behavior along the airfoil's surface. In the case of the AFC implementation, it is observed that the SJ generates a sudden rise of the friction coefficient and this prevents form becoming negative at any point along the chord of the airfoil, which suggests that the flow remains attached throughout the entire airfoil. In contrast, in the baseline case, flow separation is clearly observed at $x/c=0.423$.

\begin{figure}
    \centering
    \begin{subfigure}[t]{0.55\textwidth}
        \centering
        \includegraphics[width=\textwidth]{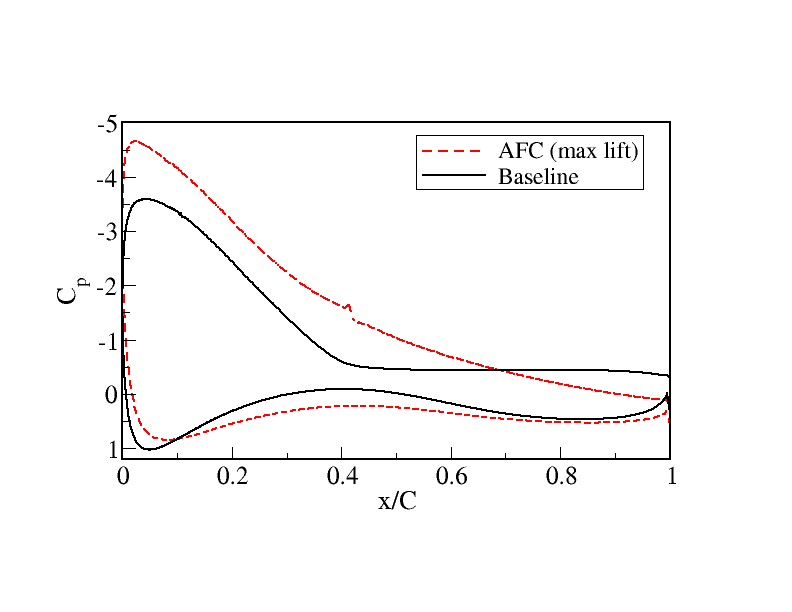}
        \caption{}
        \label{subfig:CpS54AFC}
    \end{subfigure}%
    \begin{subfigure}[t]{0.55\textwidth}
        \centering
        \includegraphics[width=\textwidth]{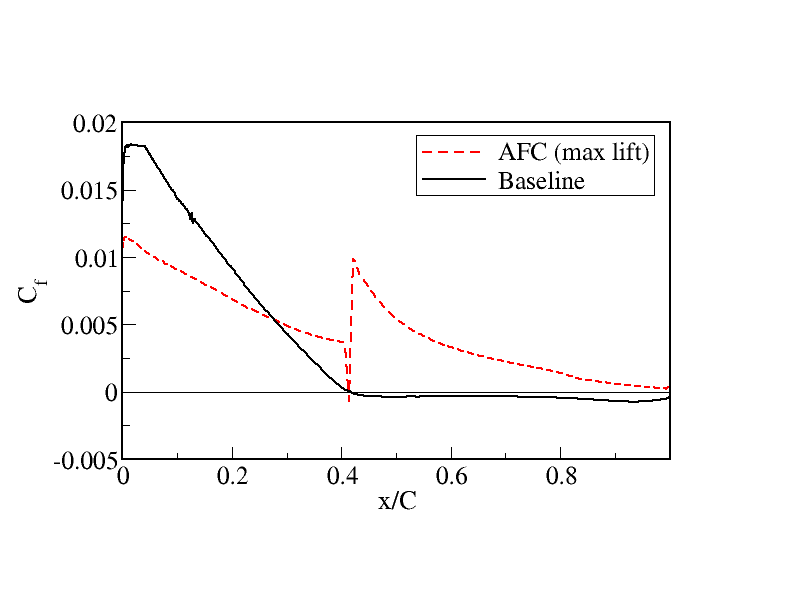}
        \caption{}
        \label{subfig:CfS54AFC}
    \end{subfigure}
    \caption{Pressure and tangential wall shear stress coefficients between the baseline case and the optimal AFC implementation on Section 54.}
    \label{fig:cpcf_Comparision}
\end{figure}

Figure \ref{velocity_profileAFC} provides a comparison between the time averaged velocity profiles tangent to the airfoil for both the AFC (actuated) and baseline cases. Regarding the AFC case, it is observed that the velocity distribution is positive at all points, therefore matching the profile of a non-separated flow. Moreover, the increase in velocity near the jet injection region is clearly visible, indicating the impact of the synthetic jet on the external flow. On the other hand, the baseline case exhibits a negative velocity distribution pattern from $x/c=0.423$ to $x/c=1$, indicating the occurrence of flow separation in all this zone.

\begin{figure}
    \centering
    \begin{subfigure}[t]{0.5\textwidth}
        \centering
        \includegraphics[width=\textwidth]{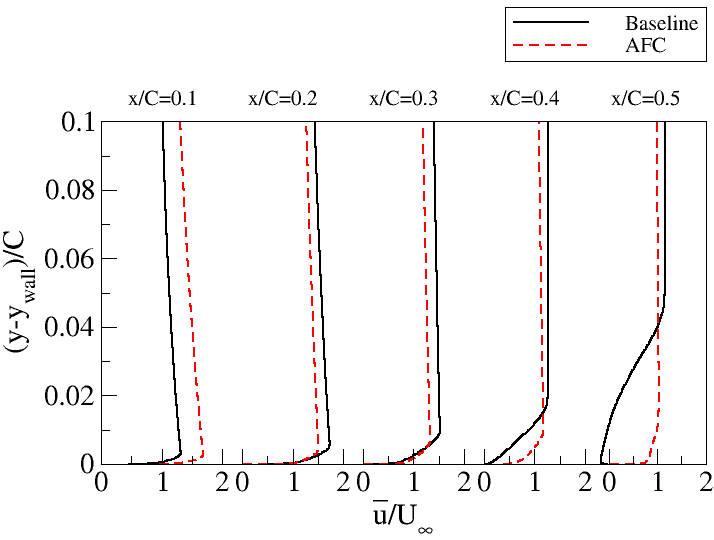}
    \end{subfigure}%
    \begin{subfigure}[t]{0.5\textwidth}
        \centering
        \includegraphics[width=\textwidth]{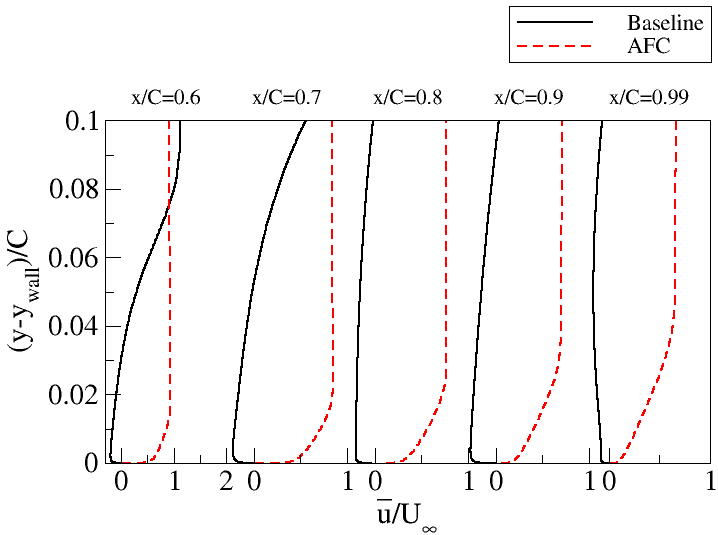}
    \end{subfigure}
     \caption{Mean velocity profiles tangent to the airfoil surface at several streamwise locations, from $x/C=0.1$ to $x/C=0.99$, and as a function of the normalised wall normal distance. Comparison of the baseline with he optimum AFC case.}
    \label{velocity_profileAFC}
\end{figure}

\subsection{Energy assessment}\label{sec:flowenergy}

In the present section it will be determined the energy balance of the AFC implementation on Section 54, by comparing the power required to delay the boundary layer separation and the power increase obtained by the wind turbine due to the AFC implementation.
The power required to activate the synthetic jet is defined as \cite{Tousi2021,de2015comparison}:
\begin{equation}
     W_j=\frac{1}{2}\rho_jA_j\sin\theta_{\text{jet}}\overline{u^3_j}
\end{equation}

Where $\rho_j$ states for the density of the exiting fluid (in this case air at sea level ISA conditions, $\rho_j=\rho_\infty=1.225\:kg/m^3$), $A_j$ is the cross-sectional flow area of the synthetic jets ($A_j=h\cdot \text{extrusion length}=h\cdot 1=0.03045\:m$) and $\overline{u^3_j}$ is the jet time dependent velocity profile defined as \cite{de2015comparison,de2012active}:

\begin{equation}
    \overline{u^3_j}=\frac{1}{T/2}\int_{0}^{T/2}u_{\max}\sin^3(2\pi ft)\mathrm{d}t=\frac{4}{3\pi}u_{\max}^3
\end{equation}

T being the SJ oscillation period.


The incremental torque represents the difference between the torque generated by the actuated section and the baseline one. It is initially essential to determine the resultant aerodynamic force ($F_{rot}$) for both cases, which is obtained by decomposing the lift and drag forces in the plane of rotation using the inflow angle ($\varphi$). It is important to note that the provided aerodynamic coefficients are time-averaged values, ensuring a representative analysis of the aerodynamic forces.
\begin{equation}
    F_{rot\:AFC}=\frac{1}{2}\rho_\infty u_{rel}^2c\left ( c_{l\:AFC}\sin\varphi- c_{d\:AFC}\cos\varphi\right ) 
\end{equation}
\begin{equation}
    F_{rot\:BSC}=\frac{1}{2}\rho_\infty u_{rel}^2c\left ( c_{l\:BCS}\sin\varphi- c_{d\:BCS}\cos\varphi\right ) 
\end{equation}
By considering the forces generated in both the actuated and baseline cases, the increase in torque $\Delta Q$ can be determined. 

\begin{equation}
    \Delta Q = Q_{AFC}-Q_{BCS} =z_{S54}\cdot \left ( F_{rot\:AFC}-F_{rot\:BCS} \right )
\end{equation}
$z_{S54}$ represents the $z$ coordinate of the improved section, its value being $31.42\:m$.

At this stage, the increase in output power resulting from the lift and drag improvement can be expressed as follows, $\Omega$ denoting the angular velocity in radians per second ($rad/s$).

\begin{equation}
    \Delta W=\Delta Q \cdot \Omega
\end{equation}
The net power balance ($W_G$) between the incremental power of the DTU 10MW RWT ($\Delta P$) and the power needed to delay the boundary layer separation of Section 54 ($W_j$) is defined as follows:
\begin{equation}
    W_G=\Delta P -W_j
    \label{eq:WG}
\end{equation}
The numerical values for the previous equation give:
\begin{equation}
   W_G = 27.982 - 1.794 = 26.188\:kW
\end{equation}
If $W_G$ becomes negative, it means that the AFC implementation could improve the lift/aerodynamic efficiency of the section but without a net power saving to the wind turbine. In contrast, whenever the power balance becomes positive, there is a gain in total energy and enhanced aerodynamic capabilities of the mentioned section.
In the current case, the power required by the synthetic jet to fully attach the flow to the airfoil, represents $6.41\:\%$ of the power gained resulting from the increased lift and reduced drag achieved through the AFC implementation. Therefore, it can be concluded that the AFC implementation via Synthetic Jets on Section 54 (DTU 10MW RWT) at a wind speed of 10 m/s, greatly increases the power output of the wind turbine.

\section{Conclusions}   \label{sec:Conclusions}

This paper clarifies that to be able to improve the performance/efficiency/power generated in any wind turbine, it is essential to analyze the flow around each airfoil including not just the aerodynamic forces (which is what has been done until now) but also the boundary layer separation point and the vortex shedding frequency/ies associated. Once this first step is accomplished the flow can be reattached in any section via using AFC.  
The present research presents a parametric analysis with the aim to optimize three AFC parameters, momentum coefficient, jet inclination angle and pulsating frequency, associated to a SJA and when applied to a wind turbine section. The procedure followed to perform a parametric optimization of any airfoil is established and presented. A maximum airfoil efficiency increase versus the baseline case of $264\%$ is obtained for $C_{\mu}$ = 0.007; $F^+$ = 3 and $\theta_{jet}$ = $5^{\circ}$ and when the groove was located at $x/C$ = 0.4269 being its width of $0.005C$. The paper further proves that the AFC implementation, in a single airfoil section so far and thanks to the parametric optimization performed, requires about $6.4\%$ of the energy increase experienced by the turbine due to the implementation of the AFC technique outlined in this paper. The peak to peak amplitude reduction of the time dependent lift and drag coefficients, versus the baselinecase, clearly indicates that the fluctuation forces on the blade can be drastically decreased when AFC is applied.

\section*{Declarations} The authors have no conflicts to disclose.
\subsection*{Ethical Approval} Not applicable.
\subsection*{Funding} This research was supported by the Universitat Politècnica de Catalunya under the grant OBLEA-2024, and by the Spanish Ministerio de Ciencia, Innovacion y Universidades with the project PID2023-150014OB-C21. Some of the computations were performed in the Red Española de Supercomputación (RES), Spanish supercomputer network, under the grant IM-2024-2-0008.

\subsection*{Availability of data and materials} Data shall be made available upon reasonable request. 

\bibliography{sn-article_1}


\begin{thebibliography}{57}
\ifx \bisbn   \undefined \def \bisbn  #1{ISBN #1}\fi
\ifx \binits  \undefined \def \binits#1{#1}\fi
\ifx \bauthor  \undefined \def \bauthor#1{#1}\fi
\ifx \batitle  \undefined \def \batitle#1{#1}\fi
\ifx \bjtitle  \undefined \def \bjtitle#1{#1}\fi
\ifx \bvolume  \undefined \def \bvolume#1{\textbf{#1}}\fi
\ifx \byear  \undefined \def \byear#1{#1}\fi
\ifx \bissue  \undefined \def \bissue#1{#1}\fi
\ifx \bfpage  \undefined \def \bfpage#1{#1}\fi
\ifx \blpage  \undefined \def \blpage #1{#1}\fi
\ifx \burl  \undefined \def \burl#1{\textsf{#1}}\fi
\ifx \doiurl  \undefined \def \doiurl#1{\url{https://doi.org/#1}}\fi
\ifx \betal  \undefined \def \betal{\textit{et al.}}\fi
\ifx \binstitute  \undefined \def \binstitute#1{#1}\fi
\ifx \binstitutionaled  \undefined \def \binstitutionaled#1{#1}\fi
\ifx \bctitle  \undefined \def \bctitle#1{#1}\fi
\ifx \beditor  \undefined \def \beditor#1{#1}\fi
\ifx \bpublisher  \undefined \def \bpublisher#1{#1}\fi
\ifx \bbtitle  \undefined \def \bbtitle#1{#1}\fi
\ifx \bedition  \undefined \def \bedition#1{#1}\fi
\ifx \bseriesno  \undefined \def \bseriesno#1{#1}\fi
\ifx \blocation  \undefined \def \blocation#1{#1}\fi
\ifx \bsertitle  \undefined \def \bsertitle#1{#1}\fi
\ifx \bsnm \undefined \def \bsnm#1{#1}\fi
\ifx \bsuffix \undefined \def \bsuffix#1{#1}\fi
\ifx \bparticle \undefined \def \bparticle#1{#1}\fi
\ifx \barticle \undefined \def \barticle#1{#1}\fi
\bibcommenthead
\ifx \bconfdate \undefined \def \bconfdate #1{#1}\fi
\ifx \botherref \undefined \def \botherref #1{#1}\fi
\ifx \url \undefined \def \url#1{\textsf{#1}}\fi
\ifx \bchapter \undefined \def \bchapter#1{#1}\fi
\ifx \bbook \undefined \def \bbook#1{#1}\fi
\ifx \bcomment \undefined \def \bcomment#1{#1}\fi
\ifx \oauthor \undefined \def \oauthor#1{#1}\fi
\ifx \citeauthoryear \undefined \def \citeauthoryear#1{#1}\fi
\ifx \endbibitem  \undefined \def \endbibitem {}\fi
\ifx \bconflocation  \undefined \def \bconflocation#1{#1}\fi
\ifx \arxivurl  \undefined \def \arxivurl#1{\textsf{#1}}\fi
\csname PreBibitemsHook\endcsname

\bibitem[\protect\citeauthoryear{Zhu et~al.}{2018}]{zhu2018simulation}
\begin{barticle}
\bauthor{\bsnm{Zhu}, \binits{H.}},
\bauthor{\bsnm{Hao}, \binits{W.}},
\bauthor{\bsnm{Li}, \binits{C.}},
\bauthor{\bsnm{Ding}, \binits{Q.}}:
\batitle{Simulation on flow control strategy of synthetic jet in an vertical axis wind turbine}.
\bjtitle{Aerospace Science and Technology}
\bvolume{77},
\bfpage{439}--\blpage{448}
(\byear{2018})
\end{barticle}
\endbibitem

\bibitem[\protect\citeauthoryear{Hochh{\"a}usler and Erfort}{2021}]{hochhausler2021experimental}
\begin{botherref}
\oauthor{\bsnm{Hochh{\"a}usler}, \binits{D.}},
\oauthor{\bsnm{Erfort}, \binits{G.}}:
Experimental study on an airfoil equipped with an active flow control element.
Wind Engineering,
0309524--211066782
(2021)
\end{botherref}
\endbibitem

\bibitem[\protect\citeauthoryear{Khalil et~al.}{2021}]{khalil2021active}
\begin{botherref}
\oauthor{\bsnm{Khalil}, \binits{K.}},
\oauthor{\bsnm{Asaro}, \binits{S.}},
\oauthor{\bsnm{Bauknecht}, \binits{A.}}:
Active flow control devices for wing load alleviation.
Journal of Aircraft,
1--17
(2021)
\end{botherref}
\endbibitem

\bibitem[\protect\citeauthoryear{Mosca et~al.}{2021}]{mosca2021multidisciplinary}
\begin{botherref}
\oauthor{\bsnm{Mosca}, \binits{V.}},
\oauthor{\bsnm{Karpuk}, \binits{S.}},
\oauthor{\bsnm{Sudhi}, \binits{A.}},
\oauthor{\bsnm{Badrya}, \binits{C.}},
\oauthor{\bsnm{Elham}, \binits{A.}}:
Multidisciplinary design optimisation of a fully electric regional aircraft wing with active flow control technology.
The Aeronautical Journal,
1--25
(2021)
\end{botherref}
\endbibitem

\bibitem[\protect\citeauthoryear{Maldonado et~al.}{2023}]{maldonado2023effect}
\begin{barticle}
\bauthor{\bsnm{Maldonado}, \binits{V.}},
\bauthor{\bsnm{Peralta}, \binits{N.}},
\bauthor{\bsnm{Ayele}, \binits{W.}},
\bauthor{\bsnm{Santos}, \binits{D.}},
\bauthor{\bsnm{Dufflis}, \binits{G.}}:
\batitle{Effect of blade aspect ratio on the performance tradeoff between figure of merit and bending-torsion dynamics of wind turbines with synthetic jets}.
\bjtitle{Energy Reports}
\bvolume{9},
\bfpage{4830}--\blpage{4843}
(\byear{2023})
\end{barticle}
\endbibitem

\bibitem[\protect\citeauthoryear{Cattafesta and Sheplak}{2011}]{Cattafesta_2011}
\begin{barticle}
\bauthor{\bsnm{Cattafesta}, \binits{L.N.}},
\bauthor{\bsnm{Sheplak}, \binits{M.}}:
\batitle{Actuators for active flow control}.
\bjtitle{Annual Review of Fluid Mechanics}
\bvolume{43}(\bissue{1}),
\bfpage{247}--\blpage{272}
(\byear{2011})
\doiurl{10.1146/annurev-fluid-122109-160634}
\end{barticle}
\endbibitem

\bibitem[\protect\citeauthoryear{Wang and bao Tian}{2019}]{Wang_2019}
\begin{botherref}
\oauthor{\bsnm{Wang}, \binits{L.}},
\oauthor{\bsnm{Tian}, \binits{F.-b.}}:
Numerical simulation of flow over a parallel cantilevered flag in the vicinity of a rigid wall.
Physical Review E
\textbf{99}(5)
(2019)
\doiurl{10.1103/physreve.99.053111}
\end{botherref}
\endbibitem

\bibitem[\protect\citeauthoryear{Cho and Shyy}{2011}]{Cho_2011}
\begin{barticle}
\bauthor{\bsnm{Cho}, \binits{Y.-C.}},
\bauthor{\bsnm{Shyy}, \binits{W.}}:
\batitle{Adaptive flow control of low-{R}eynolds number aerodynamics using dielectric barrier discharge actuator}.
\bjtitle{Progress in Aerospace Sciences}
\bvolume{47}(\bissue{7}),
\bfpage{495}--\blpage{521}
(\byear{2011})
\doiurl{10.1016/j.paerosci.2011.06.005}
\end{barticle}
\endbibitem

\bibitem[\protect\citeauthoryear{Foshat}{2020}]{Foshat_2020}
\begin{barticle}
\bauthor{\bsnm{Foshat}, \binits{S.}}:
\batitle{Numerical investigation of the effects of plasma actuator on separated laminar flows past an incident plate under ground effect}.
\bjtitle{Aerospace Science and Technology}
\bvolume{98},
\bfpage{105646}
(\byear{2020})
\doiurl{10.1016/j.ast.2019.105646}
\end{barticle}
\endbibitem

\bibitem[\protect\citeauthoryear{Benard and Moreau}{2014}]{Benard_2014}
\begin{botherref}
\oauthor{\bsnm{Benard}, \binits{N.}},
\oauthor{\bsnm{Moreau}, \binits{E.}}:
Electrical and mechanical characteristics of surface {AC} dielectric barrier discharge plasma actuators applied to airflow control.
Experiments in Fluids
\textbf{55}(11)
(2014)
\doiurl{10.1007/s00348-014-1846-x}
\end{botherref}
\endbibitem

\bibitem[\protect\citeauthoryear{Benard et~al.}{2016}]{Benard_2016}
\begin{botherref}
\oauthor{\bsnm{Benard}, \binits{N.}},
\oauthor{\bsnm{Pons-Prats}, \binits{J.}},
\oauthor{\bsnm{Periaux}, \binits{J.}},
\oauthor{\bsnm{Bugeda}, \binits{G.}},
\oauthor{\bsnm{Braud}, \binits{P.}},
\oauthor{\bsnm{Bonnet}, \binits{J.}},
\oauthor{\bsnm{Moreau}, \binits{E.}}:
Turbulent separated shear flow control by surface plasma actuator: experimental optimization by genetic algorithm approach.
Experiments in Fluids
\textbf{57}(2)
(2016)
\doiurl{10.1007/s00348-015-2107-3}
\end{botherref}
\endbibitem

\bibitem[\protect\citeauthoryear{Bergad{\`a} et~al.}{2021}]{bergada2021fluidic}
\begin{barticle}
\bauthor{\bsnm{Bergad{\`a}}, \binits{J.M.}},
\bauthor{\bsnm{Baghaei}, \binits{M.}},
\bauthor{\bsnm{Prakash}, \binits{B.}},
\bauthor{\bsnm{Mellibovsky}, \binits{F.}}:
\batitle{Fluidic oscillators, feedback channel effect under compressible flow conditions}.
\bjtitle{Sensors}
\bvolume{21}(\bissue{17}),
\bfpage{5768}
(\byear{2021})
\end{barticle}
\endbibitem

\bibitem[\protect\citeauthoryear{Baghaei and Bergada}{2020}]{Baghaei_2020}
\begin{barticle}
\bauthor{\bsnm{Baghaei}, \binits{M.}},
\bauthor{\bsnm{Bergada}, \binits{J.M.}}:
\batitle{Fluidic oscillators, the effect of some design modifications}.
\bjtitle{Applied Sciences}
\bvolume{10}(\bissue{6}),
\bfpage{2105}
(\byear{2020})
\doiurl{10.3390/app10062105}
\end{barticle}
\endbibitem

\bibitem[\protect\citeauthoryear{Baghaei and Bergada}{2019}]{Baghaei_2019}
\begin{barticle}
\bauthor{\bsnm{Baghaei}, \binits{M.}},
\bauthor{\bsnm{Bergada}, \binits{J.M.}}:
\batitle{Analysis of the forces driving the oscillations in 3d fluidic oscillators}.
\bjtitle{Energies}
\bvolume{12}(\bissue{24}),
\bfpage{4720}
(\byear{2019})
\doiurl{10.3390/en12244720}
\end{barticle}
\endbibitem

\bibitem[\protect\citeauthoryear{Glezer and Amitay}{2002}]{glezer2002synthetic}
\begin{barticle}
\bauthor{\bsnm{Glezer}, \binits{A.}},
\bauthor{\bsnm{Amitay}, \binits{M.}}:
\batitle{Synthetic jets}.
\bjtitle{Annual review of fluid mechanics}
\bvolume{34}(\bissue{1}),
\bfpage{503}--\blpage{529}
(\byear{2002})
\end{barticle}
\endbibitem

\bibitem[\protect\citeauthoryear{Rumsey et~al.}{2004}]{rumsey2004summary}
\begin{bchapter}
\bauthor{\bsnm{Rumsey}, \binits{C.}},
\bauthor{\bsnm{Gatski}, \binits{T.}},
\bauthor{\bsnm{Sellers}, \binits{W.}},
\bauthor{\bsnm{Vatsa}, \binits{V.}},
\bauthor{\bsnm{Viken}, \binits{S.}}:
\bctitle{Summary of the 2004 cfd validation workshop on synthetic jets and turbulent separation control}.
In: \bbtitle{2nd AIAA Flow Control Conference},
p. \bfpage{2217}
(\byear{2004})
\end{bchapter}
\endbibitem

\bibitem[\protect\citeauthoryear{Wygnanski}{2004}]{wygnanski2004variables}
\begin{bchapter}
\bauthor{\bsnm{Wygnanski}, \binits{I.}}:
\bctitle{The variables affecting the control of separation by periodic excitation}.
In: \bbtitle{2nd AIAA Flow Control Conference},
p. \bfpage{2505}
(\byear{2004})
\end{bchapter}
\endbibitem

\bibitem[\protect\citeauthoryear{Findanis and Ahmed}{2008}]{findanis2008interaction}
\begin{barticle}
\bauthor{\bsnm{Findanis}, \binits{N.}},
\bauthor{\bsnm{Ahmed}, \binits{N.}}:
\batitle{The interaction of an asymmetrical localised synthetic jet on a side-supported sphere}.
\bjtitle{Journal of Fluids and Structures}
\bvolume{24}(\bissue{7}),
\bfpage{1006}--\blpage{1020}
(\byear{2008})
\doiurl{10.1016/j.jfluidstructs.2008.02.002}
\end{barticle}
\endbibitem

\bibitem[\protect\citeauthoryear{De~Giorgi et~al.}{2015}]{de2015comparison}
\begin{barticle}
\bauthor{\bsnm{De~Giorgi}, \binits{M.}},
\bauthor{\bsnm{De~Luca}, \binits{C.}},
\bauthor{\bsnm{Ficarella}, \binits{A.}},
\bauthor{\bsnm{Marra}, \binits{F.}}:
\batitle{Comparison between synthetic jets and continuous jets for active flow control: Application on a {NACA} 0015 and a compressor stator cascade}.
\bjtitle{Aerospace Science and Technology}
\bvolume{43},
\bfpage{256}--\blpage{280}
(\byear{2015})
\doiurl{10.1016/j.ast.2015.03.004}
\end{barticle}
\endbibitem

\bibitem[\protect\citeauthoryear{Traficante et~al.}{2016}]{traficante2016flow}
\begin{barticle}
\bauthor{\bsnm{Traficante}, \binits{S.}},
\bauthor{\bsnm{De~Giorgi}, \binits{M.}},
\bauthor{\bsnm{Ficarella}, \binits{A.}}:
\batitle{Flow separation control on a compressor-stator cascade using plasma actuators and synthetic and continuous jets}.
\bjtitle{Journal of Aerospace Engineering}
\bvolume{29}(\bissue{3}),
\bfpage{04015056}
(\byear{2016})
\end{barticle}
\endbibitem

\bibitem[\protect\citeauthoryear{Zhang et~al.}{2019}]{zhang2019comparison}
\begin{barticle}
\bauthor{\bsnm{Zhang}, \binits{H.}},
\bauthor{\bsnm{Chen}, \binits{S.}},
\bauthor{\bsnm{Gong}, \binits{Y.}},
\bauthor{\bsnm{Wang}, \binits{S.}}:
\batitle{A comparison of different unsteady flow control techniques in a highly loaded compressor cascade}.
\bjtitle{Proceedings of the Institution of Mechanical Engineers, Part G: Journal of Aerospace Engineering}
\bvolume{233}(\bissue{6}),
\bfpage{2051}--\blpage{2065}
(\byear{2019})
\end{barticle}
\endbibitem

\bibitem[\protect\citeauthoryear{Tousi et~al.}{2021}]{Tousi2021}
\begin{barticle}
\bauthor{\bsnm{Tousi}, \binits{N.M.}},
\bauthor{\bsnm{Coma}, \binits{M.}},
\bauthor{\bsnm{Bergada}, \binits{J.M.}},
\bauthor{\bsnm{Pons-Prats}, \binits{J.}},
\bauthor{\bsnm{Mellibovsky}, \binits{F.}},
\bauthor{\bsnm{Bugeda}, \binits{G.}}:
\batitle{Active flow control optimisation on sd7003 airfoil at pre and post-stall angles of attack using synthetic jets}.
\bjtitle{Applied Mathematical Modelling}
(\byear{2021})
\doiurl{10.1016/j.apm.2021.05.016}
\end{barticle}
\endbibitem

\bibitem[\protect\citeauthoryear{Tadjfar and Kamari}{2020}]{tadjfar2020optimization}
\begin{botherref}
\oauthor{\bsnm{Tadjfar}, \binits{M.}},
\oauthor{\bsnm{Kamari}, \binits{D.}}:
Optimization of flow control parameters over {SD}7003 airfoil with synthetic jet actuator.
Journal of Fluids Engineering
\textbf{142}(2)
(2020)
\doiurl{10.1115/1.4044985}
\end{botherref}
\endbibitem

\bibitem[\protect\citeauthoryear{Tousi et~al.}{2022}]{tousi2022large}
\begin{barticle}
\bauthor{\bsnm{Tousi}, \binits{N.}},
\bauthor{\bsnm{Bergad{\`a}}, \binits{J.}},
\bauthor{\bsnm{Mellibovsky}, \binits{F.}}:
\batitle{Large eddy simulation of optimal synthetic jet actuation on a sd7003 airfoil in post-stall conditions}.
\bjtitle{Aerospace Science and Technology}
\bvolume{127},
\bfpage{107679}
(\byear{2022})
\end{barticle}
\endbibitem

\bibitem[\protect\citeauthoryear{Stalnov et~al.}{2010}]{stalnov2010evaluation}
\begin{botherref}
\oauthor{\bsnm{Stalnov}, \binits{O.}},
\oauthor{\bsnm{Kribus}, \binits{A.}},
\oauthor{\bsnm{Seifert}, \binits{A.}}:
Evaluation of active flow control applied to wind turbine blade section.
Journal of Renewable and Sustainable Energy
\textbf{2}(6)
(2010)
\end{botherref}
\endbibitem

\bibitem[\protect\citeauthoryear{Maldonado et~al.}{2010}]{maldonado2010active}
\begin{barticle}
\bauthor{\bsnm{Maldonado}, \binits{V.}},
\bauthor{\bsnm{Farnsworth}, \binits{J.}},
\bauthor{\bsnm{Gressick}, \binits{W.}},
\bauthor{\bsnm{Amitay}, \binits{M.}}:
\batitle{Active control of flow separation and structural vibrations of wind turbine blades}.
\bjtitle{Wind Energy: An International Journal for Progress and Applications in Wind Power Conversion Technology}
\bvolume{13}(\bissue{2-3}),
\bfpage{221}--\blpage{237}
(\byear{2010})
\end{barticle}
\endbibitem

\bibitem[\protect\citeauthoryear{Yen and Ahmed}{2013}]{yen2013enhancing}
\begin{barticle}
\bauthor{\bsnm{Yen}, \binits{J.}},
\bauthor{\bsnm{Ahmed}, \binits{N.A.}}:
\batitle{Enhancing vertical axis wind turbine by dynamic stall control using synthetic jets}.
\bjtitle{Journal of Wind Engineering and Industrial Aerodynamics}
\bvolume{114},
\bfpage{12}--\blpage{17}
(\byear{2013})
\end{barticle}
\endbibitem

\bibitem[\protect\citeauthoryear{Taylor et~al.}{2015}]{taylor2015load}
\begin{barticle}
\bauthor{\bsnm{Taylor}, \binits{K.}},
\bauthor{\bsnm{Leong}, \binits{C.M.}},
\bauthor{\bsnm{Amitay}, \binits{M.}}:
\batitle{Load control on a dynamically pitching finite span wind turbine blade using synthetic jets}.
\bjtitle{Wind Energy}
\bvolume{18}(\bissue{10}),
\bfpage{1759}--\blpage{1775}
(\byear{2015})
\end{barticle}
\endbibitem

\bibitem[\protect\citeauthoryear{Velasco et~al.}{2017}]{velasco2017numerical}
\begin{barticle}
\bauthor{\bsnm{Velasco}, \binits{D.}},
\bauthor{\bsnm{Mejia}, \binits{O.L.}},
\bauthor{\bsnm{La{\'\i}n}, \binits{S.}}:
\batitle{Numerical simulations of active flow control with synthetic jets in a darrieus turbine}.
\bjtitle{Renewable Energy}
\bvolume{113},
\bfpage{129}--\blpage{140}
(\byear{2017})
\end{barticle}
\endbibitem

\bibitem[\protect\citeauthoryear{Moshfeghi and Hur}{2017}]{moshfeghi2017numerical}
\begin{barticle}
\bauthor{\bsnm{Moshfeghi}, \binits{M.}},
\bauthor{\bsnm{Hur}, \binits{N.}}:
\batitle{Numerical study on the effects of a synthetic jet actuator on s809 airfoil aerodynamics at different flow regimes and jet flow angles}.
\bjtitle{Journal of Mechanical Science and Technology}
\bvolume{31},
\bfpage{1233}--\blpage{1240}
(\byear{2017})
\end{barticle}
\endbibitem

\bibitem[\protect\citeauthoryear{Maldonado and Gupta}{2019}]{maldonado2019increasing}
\begin{barticle}
\bauthor{\bsnm{Maldonado}, \binits{V.}},
\bauthor{\bsnm{Gupta}, \binits{S.}}:
\batitle{Increasing the power efficiency of rotors at transitional reynolds numbers with synthetic jet actuators}.
\bjtitle{Experimental Thermal and Fluid Science}
\bvolume{105},
\bfpage{356}--\blpage{366}
(\byear{2019})
\end{barticle}
\endbibitem

\bibitem[\protect\citeauthoryear{Wang et~al.}{2022}]{wang2022effect}
\begin{barticle}
\bauthor{\bsnm{Wang}, \binits{P.}},
\bauthor{\bsnm{Liu}, \binits{Q.}},
\bauthor{\bsnm{Li}, \binits{C.}},
\bauthor{\bsnm{Miao}, \binits{W.}},
\bauthor{\bsnm{Luo}, \binits{S.}},
\bauthor{\bsnm{Sun}, \binits{K.}},
\bauthor{\bsnm{Niu}, \binits{K.}}:
\batitle{Effect of trailing edge dual synthesis jets actuator on aerodynamic characteristics of a straight-bladed vertical axis wind turbine}.
\bjtitle{Energy}
\bvolume{238},
\bfpage{121792}
(\byear{2022})
\end{barticle}
\endbibitem

\bibitem[\protect\citeauthoryear{Maldonado et~al.}{2009}]{maldonado2009active}
\begin{botherref}
\oauthor{\bsnm{Maldonado}, \binits{V.}},
\oauthor{\bsnm{Boucher}, \binits{M.}},
\oauthor{\bsnm{Ostman}, \binits{R.}},
\oauthor{\bsnm{Amitay}, \binits{M.}}:
Active vibration control of a wind turbine blade using synthetic jets.
International Journal of Flow Control
\textbf{1}(4)
(2009)
\end{botherref}
\endbibitem

\bibitem[\protect\citeauthoryear{Goodfellow et~al.}{2013}]{goodfellow2013momentum}
\begin{barticle}
\bauthor{\bsnm{Goodfellow}, \binits{S.D.}},
\bauthor{\bsnm{Yarusevych}, \binits{S.}},
\bauthor{\bsnm{Sullivan}, \binits{P.E.}}:
\batitle{Momentum coefficient as a parameter for aerodynamic flow control with synthetic jets}.
\bjtitle{{AIAA} Journal}
\bvolume{51}(\bissue{3}),
\bfpage{623}--\blpage{631}
(\byear{2013})
\doiurl{10.2514/1.j051935}
\end{barticle}
\endbibitem

\bibitem[\protect\citeauthoryear{Feero et~al.}{2017}]{feero2017influence}
\begin{botherref}
\oauthor{\bsnm{Feero}, \binits{M.A.}},
\oauthor{\bsnm{Lavoie}, \binits{P.}},
\oauthor{\bsnm{Sullivan}, \binits{P.E.}}:
Influence of synthetic jet location on active control of an airfoil at low {R}eynolds number.
Experiments in Fluids
\textbf{58}(8)
(2017)
\doiurl{10.1007/s00348-017-2387-x}
\end{botherref}
\endbibitem

\bibitem[\protect\citeauthoryear{Amitay et~al.}{2001}]{amitay2001aerodynamic}
\begin{barticle}
\bauthor{\bsnm{Amitay}, \binits{M.}},
\bauthor{\bsnm{Smith}, \binits{D.R.}},
\bauthor{\bsnm{Kibens}, \binits{V.}},
\bauthor{\bsnm{Parekh}, \binits{D.E.}},
\bauthor{\bsnm{Glezer}, \binits{A.}}:
\batitle{Aerodynamic flow control over an unconventional airfoil using synthetic jet actuators}.
\bjtitle{{AIAA} Journal}
\bvolume{39},
\bfpage{361}--\blpage{370}
(\byear{2001})
\doiurl{10.2514/3.14740}
\end{barticle}
\endbibitem

\bibitem[\protect\citeauthoryear{Amitay and Glezer}{2002}]{amitay2002role}
\begin{barticle}
\bauthor{\bsnm{Amitay}, \binits{M.}},
\bauthor{\bsnm{Glezer}, \binits{A.}}:
\batitle{Role of actuation frequency in controlled flow reattachment over a stalled airfoil}.
\bjtitle{{AIAA} Journal}
\bvolume{40},
\bfpage{209}--\blpage{216}
(\byear{2002})
\doiurl{10.2514/3.15052}
\end{barticle}
\endbibitem

\bibitem[\protect\citeauthoryear{Feero et~al.}{2015}]{feero2015flow}
\begin{barticle}
\bauthor{\bsnm{Feero}, \binits{M.A.}},
\bauthor{\bsnm{Goodfellow}, \binits{S.D.}},
\bauthor{\bsnm{Lavoie}, \binits{P.}},
\bauthor{\bsnm{Sullivan}, \binits{P.E.}}:
\batitle{Flow reattachment using synthetic jet actuation on a low-{R}eynolds-number airfoil}.
\bjtitle{{AIAA} Journal}
\bvolume{53}(\bissue{7}),
\bfpage{2005}--\blpage{2014}
(\byear{2015})
\doiurl{10.2514/1.j053605}
\end{barticle}
\endbibitem

\bibitem[\protect\citeauthoryear{Tuck and Soria}{2008}]{tuck2008separation}
\begin{barticle}
\bauthor{\bsnm{Tuck}, \binits{A.}},
\bauthor{\bsnm{Soria}, \binits{J.}}:
\batitle{Separation control on a {NACA} 0015 airfoil using a 2d micro {ZNMF} jet}.
\bjtitle{Aircraft Engineering and Aerospace Technology}
\bvolume{80}(\bissue{2}),
\bfpage{175}--\blpage{180}
(\byear{2008})
\doiurl{10.1108/00022660810859391}
\end{barticle}
\endbibitem

\bibitem[\protect\citeauthoryear{Kitsios et~al.}{2011}]{kitsios2011coherent}
\begin{barticle}
\bauthor{\bsnm{Kitsios}, \binits{V.}},
\bauthor{\bsnm{Cordier}, \binits{L.}},
\bauthor{\bsnm{Bonnet}, \binits{J.-P.}},
\bauthor{\bsnm{Ooi}, \binits{A.}},
\bauthor{\bsnm{Soria}, \binits{J.}}:
\batitle{On the coherent structures and stability properties of a leading-edge separated aerofoil with turbulent recirculation}.
\bjtitle{Journal of Fluid Mechanics}
\bvolume{683},
\bfpage{395}--\blpage{416}
(\byear{2011})
\doiurl{10.1017/jfm.2011.285}
\end{barticle}
\endbibitem

\bibitem[\protect\citeauthoryear{Buchmann et~al.}{2013}]{buchmann2013influence}
\begin{botherref}
\oauthor{\bsnm{Buchmann}, \binits{N.}},
\oauthor{\bsnm{Atkinson}, \binits{C.}},
\oauthor{\bsnm{Soria}, \binits{J.}}:
Influence of {ZNMF} jet flow control on the spatio-temporal flow structure over a {NACA}-0015 airfoil.
Experiments in Fluids
\textbf{54}(3)
(2013)
\doiurl{10.1007/s00348-013-1485-7}
\end{botherref}
\endbibitem

\bibitem[\protect\citeauthoryear{Zhang and Samtaney}{2015}]{zhang2015direct}
\begin{barticle}
\bauthor{\bsnm{Zhang}, \binits{W.}},
\bauthor{\bsnm{Samtaney}, \binits{R.}}:
\batitle{A direct numerical simulation investigation of the synthetic jet frequency effects on separation control of low-{R}e flow past an airfoil}.
\bjtitle{Physics of Fluids}
\bvolume{27}(\bissue{5}),
\bfpage{055101}
(\byear{2015})
\doiurl{10.1063/1.4919599}
\end{barticle}
\endbibitem

\bibitem[\protect\citeauthoryear{Kim and Kim}{2009}]{kim2009separation}
\begin{barticle}
\bauthor{\bsnm{Kim}, \binits{S.H.}},
\bauthor{\bsnm{Kim}, \binits{C.}}:
\batitle{Separation control on {NACA}23012 using synthetic jet}.
\bjtitle{Aerospace Science and Technology}
\bvolume{13}(\bissue{4-5}),
\bfpage{172}--\blpage{182}
(\byear{2009})
\doiurl{10.1016/j.ast.2008.11.001}
\end{barticle}
\endbibitem

\bibitem[\protect\citeauthoryear{Monir et~al.}{2014}]{monir2014tangential}
\begin{barticle}
\bauthor{\bsnm{Monir}, \binits{H.E.}},
\bauthor{\bsnm{Tadjfar}, \binits{M.}},
\bauthor{\bsnm{Bakhtian}, \binits{A.}}:
\batitle{Tangential synthetic jets for separation control}.
\bjtitle{Journal of Fluids and Structures}
\bvolume{45},
\bfpage{50}--\blpage{65}
(\byear{2014})
\doiurl{10.1016/j.jfluidstructs.2013.11.011}
\end{barticle}
\endbibitem

\bibitem[\protect\citeauthoryear{Menter}{1993}]{Menter}
\begin{bchapter}
\bauthor{\bsnm{Menter}, \binits{F.}}:
\bctitle{Zonal two equation kw turbulence models for aerodynamic flows}.
In: \bbtitle{23rd Fluid Dynamics, Plasmadynamics, and Lasers Conference},
p. \bfpage{2906}
(\byear{1993})
\end{bchapter}
\endbibitem

\bibitem[\protect\citeauthoryear{Couto and Bergada}{2022}]{couto2022aerodynamic}
\begin{barticle}
\bauthor{\bsnm{Couto}, \binits{N.}},
\bauthor{\bsnm{Bergada}, \binits{J.M.}}:
\batitle{Aerodynamic efficiency improvement on a naca-8412 airfoil via active flow control implementation}.
\bjtitle{Applied Sciences}
\bvolume{12}(\bissue{9}),
\bfpage{4269}
(\byear{2022})
\end{barticle}
\endbibitem

\bibitem[\protect\citeauthoryear{Zahle et~al.}{2014}]{zahle2014comprehensive}
\begin{bchapter}
\bauthor{\bsnm{Zahle}, \binits{F.}},
\bauthor{\bsnm{Bak}, \binits{C.}},
\bauthor{\bsnm{S{\o}rensen}, \binits{N.N.}},
\bauthor{\bsnm{Guntur}, \binits{S.}},
\bauthor{\bsnm{Troldborg}, \binits{N.}}:
\bctitle{Comprehensive aerodynamic analysis of a 10 mw wind turbine rotor using 3d cfd}.
In: \bbtitle{32nd ASME Wind Energy Symposium},
p. \bfpage{0359}
(\byear{2014})
\end{bchapter}
\endbibitem

\bibitem[\protect\citeauthoryear{Akon et~al.}{2012}]{treballinductionfactors}
\begin{botherref}
\oauthor{\bsnm{Akon}, \binits{A.F.}},
\oauthor{\bsnm{Samner}, \binits{D.}},
\oauthor{\bsnm{Trovi}, \binits{D.}}:
Measurement of axial induction factor for a model wind turbine.
PhD thesis,
Citeseer
(2012)
\end{botherref}
\endbibitem

\bibitem[\protect\citeauthoryear{Moriarty and Hansen}{2005}]{BEMllibre}
\begin{botherref}
\oauthor{\bsnm{Moriarty}, \binits{P.J.}},
\oauthor{\bsnm{Hansen}, \binits{A.C.}}:
Aerodyn theory manual.
Technical report,
National Renewable Energy Lab., Golden, CO (US)
(2005)
\end{botherref}
\endbibitem

\bibitem[\protect\citeauthoryear{Mahmuddin}{2017}]{BEMtreball}
\begin{barticle}
\bauthor{\bsnm{Mahmuddin}, \binits{F.}}:
\batitle{Rotor blade performance analysis with blade element momentum theory}.
\bjtitle{Energy Procedia}
\bvolume{105},
\bfpage{1123}--\blpage{1129}
(\byear{2017})
\end{barticle}
\endbibitem

\bibitem[\protect\citeauthoryear{Bak et~al.}{2013}]{web:DTUReport}
\begin{bchapter}
\bauthor{\bsnm{Bak}, \binits{C.}},
\bauthor{\bsnm{Zahle}, \binits{F.}},
\bauthor{\bsnm{Bitsche}, \binits{R.}},
\bauthor{\bsnm{Kim}, \binits{T.}},
\bauthor{\bsnm{Yde}, \binits{A.}},
\bauthor{\bsnm{Henriksen}, \binits{L.C.}},
\bauthor{\bsnm{Hansen}, \binits{M.H.}},
\bauthor{\bsnm{Blasques}, \binits{J.P.A.A.}},
\bauthor{\bsnm{Gaunaa}, \binits{M.}},
\bauthor{\bsnm{Natarajan}, \binits{A.}}:
\bctitle{The dtu 10-mw reference wind turbine}.
In: \bbtitle{Danish Wind Power Research 2013}
(\byear{2013})
\end{bchapter}
\endbibitem

\bibitem[\protect\citeauthoryear{Meyers}{2001}]{book:Coanda}
\begin{bbook}
\bauthor{\bsnm{Meyers}, \binits{R.A.}}:
\bbtitle{Encyclopedia of Physical Science and Technology},
\bedition{3}rd edn.
\bpublisher{Elsevier Science},
\blocation{California, United States}
(\byear{2001})
\end{bbook}
\endbibitem

\bibitem[\protect\citeauthoryear{Kim and Kim}{2009}]{momcoeff}
\begin{barticle}
\bauthor{\bsnm{Kim}, \binits{S.H.}},
\bauthor{\bsnm{Kim}, \binits{C.}}:
\batitle{Separation control on naca23012 using synthetic jet}.
\bjtitle{Aerospace Science and Technology}
\bvolume{13}(\bissue{4-5}),
\bfpage{172}--\blpage{182}
(\byear{2009})
\end{barticle}
\endbibitem

\bibitem[\protect\citeauthoryear{Tadjfar and Asgari}{2018}]{angle1}
\begin{botherref}
\oauthor{\bsnm{Tadjfar}, \binits{M.}},
\oauthor{\bsnm{Asgari}, \binits{E.}}:
The role of frequency and phase difference between the flow and the actuation signal of a tangential synthetic jet on dynamic stall flow control.
Journal of Fluids Engineering
\textbf{140}(11)
(2018)
\end{botherref}
\endbibitem

\bibitem[\protect\citeauthoryear{Krohnert}{2016}]{angle2}
\begin{bchapter}
\bauthor{\bsnm{Krohnert}, \binits{A.}}:
\bctitle{Numerical investigation of unsteady tangential blowing at the rudder of a vertical tailplane airfoil}.
In: \bbtitle{New Results in Numerical and Experimental Fluid Mechanics X: Contributions to the 19th STAB/DGLR Symposium Munich, Germany, 2014},
pp. \bfpage{39}--\blpage{49}
(\byear{2016}).
\bcomment{Springer}
\end{bchapter}
\endbibitem

\bibitem[\protect\citeauthoryear{Naim et~al.}{2007}]{angle3}
\begin{barticle}
\bauthor{\bsnm{Naim}, \binits{A.}},
\bauthor{\bsnm{Greenblatt}, \binits{D.}},
\bauthor{\bsnm{Seifert}, \binits{A.}},
\bauthor{\bsnm{Wygnanski}, \binits{I.}}:
\batitle{Active control of a circular cylinder flow at transitional reynolds numbers}.
\bjtitle{Flow, turbulence and combustion}
\bvolume{78},
\bfpage{383}--\blpage{407}
(\byear{2007})
\end{barticle}
\endbibitem

\bibitem[\protect\citeauthoryear{De~Giorgi et~al.}{2012}]{de2012active}
\begin{bchapter}
\bauthor{\bsnm{De~Giorgi}, \binits{M.G.}},
\bauthor{\bsnm{Traficante}, \binits{S.}},
\bauthor{\bsnm{De~Luca}, \binits{C.}},
\bauthor{\bsnm{Bello}, \binits{D.}},
\bauthor{\bsnm{Ficarella}, \binits{A.}}:
\bctitle{Active flow control techniques on a stator compressor cascade: a comparison between synthetic jet and plasma actuators}.
In: \bbtitle{Turbo Expo: Power for Land, Sea, and Air},
vol. \bseriesno{44748},
pp. \bfpage{439}--\blpage{450}
(\byear{2012}).
\bcomment{American Society of Mechanical Engineers}
\end{bchapter}
\endbibitem

\end{thebibliography}

\end{document}